\documentclass[journal]{IEEEtran}
\ifCLASSINFOpdf
\else
\fi
\usepackage{amsthm}
\usepackage{hyperref}
\usepackage{tikz}
\usetikzlibrary{plotmarks}
\usepackage{hyperref}

\usepackage{amsmath}
\usepackage{amssymb}
\usepackage{graphicx}
\usepackage{amsfonts}
\usepackage{latexsym}
\usepackage{pdfsync}
\usepackage{subfigure}
\usepackage{color}

 \usepackage{multirow}
  \usepackage{tabularx}
 \usepackage[english]{babel}
 \usepackage{array}

\DeclareMathAlphabet{\mathpzc}{OT1}{pzc}{m}{it}

\newcommand{\commentout}[1]{}

\newcommand{\nwc}{\newcommand}
\nwc{\ba}{\begin{array}}
\nwc{\bal}{\begin{align}}
\nwc{\bea}{\begin{eqnarray}}
\nwc{\beq}{\begin{eqnarray}}
\nwc{\bean}{\begin{eqnarray*}}
\nwc{\beqn}{\begin{eqnarray*}}
\nwc{\beqast}{\begin{eqnarray*}}

\nwc{\ea}{\end{array}}
\nwc{\eal}{\end{align}}
\nwc{\eea}{\end{eqnarray}}
\nwc{\eeq}{\end{eqnarray}}
\nwc{\eean}{\end{eqnarray*}}
\nwc{\eeqn}{\end{eqnarray*}}
\nwc{\eeqast}{\end{eqnarray*}}

\nwc{\ep}{\varepsilon}
\nwc{\ept}{\epsilon}

\newtheorem{proposition}{Proposition}

\newtheorem{theorem}{Theorem}

\newtheorem{lemma}{Lemma}
\newtheorem{remark}{Remark}

\nwc{\nn}{\nonumber}

\newcommand{\EE}{\mathbb{E}}
\newcommand{\PP}{\mathbb{P}}

\newcommand{\CC}{\mathbb{C}}
\newcommand{\RR}{\mathbb{R}}
\newcommand{\ZZ}{\mathbb{Z}}
\newcommand{\TT}{\mathbb{T}}
\newcommand{\om}{\omega}
\newcommand{\rank}{\text{rank}}
\newcommand{\supp}{\mathcal{S}}
\nwc{\calP}{\mathcal{P}}
\nwc{\calL}{\mathcal{L}}
\nwc{\itd}{\mathpzc{d}}

\nwc{\xmax}{x_{\text{max}}}
\nwc{\xmin}{x_{\text{min}}}
\nwc{\ye}{y^e}
\nwc{\range}{{\rm Range}}
\nwc{\calE}{\mathcal{E}}
\nwc{\calN}{\mathcal{N}}
\nwc{\cd}{\mathcal{d}}
\nwc{\smax}{\sigma_{\rm max}}
\nwc{\smin}{\sigma_{\rm min}}
\nwc{\re}{{\rm Re}}
\nwc{\im}{{\rm Im}}
\nwc{\bv}{{\rm v}}
\nwc{\bw}{{\rm w}}
\nwc{\PO}{\calP_1}
\nwc{\PT}{\calP_2}
\nwc{\PEO}{\calP^e_1}
\nwc{\PET}{\calP^e_2}
\nwc{\JE}{{\mathcal J}^e}
\nwc{\RE}{{\mathcal R}^e}
\nwc{\CR}{{\mathcal R}}
\nwc{\CJ}{{\mathcal J}}

\nwc{\si}{\sigma}
\nwc{\se}{\sigma^e}
\nwc{\UE}{U^e}
\nwc{\VE}{V^e}
\nwc{\lan}{\langle}
\nwc{\ran}{\rangle}

\nwc{\XP}{X(\Phi^{M-L})^T}

\nwc{\diag}{\text{diag}}
\nwc{\Hankel}{\text{Hankel}}
\nwc{\bt}{\mathbf{t}}
\nwc{\bom}{\boldsymbol\omega}
\nwc{\bN}{\mathbf{N}}
\nwc{\bn}{\mathbf{n}}
\nwc{\bm}{\mathbf{m}}
\nwc{\bL}{\mathbf{L}}
\nwc{\bc}{\mathbf{c}}
\nwc{\bH}{\mathbf{H}}
\nwc{\bh}{\mathbf{h}}
\nwc{\bG}{\mathbf{G}}
\nwc{\bg}{\mathbf{g}}
\nwc{\bz}{\mathbf{z}}
\nwc{\br}{\mathbf{r}}
\nwc{\bx}{\mathbf{x}}
\nwc{\bq}{\mathbf{q}}

\nwc{\sgn}{{\rm sgn}}
\nwc{\inn}{{[-n,n]}}
\nwc{\LLD}{{[-\frac {L_k}{2},\frac {L_k} {2}]^D}}
\nwc{\LL}{{[-\frac \bL 2,\frac \bL 2]}}
\nwc{\LkLk}{{[-\frac{L_k}{2},\frac{L_k}{2}]}}

\nwc{\hbx}{\widehat{\bx}}

\begin{document}
%
\title{MUSIC for multidimensional spectral estimation: stability and super-resolution}
%
%
%

\author{Wenjing Liao
\thanks{W. Liao is with the Department
of Mathematics, Duke University and Statistical and Applied Mathematical Sciences Institute (SAMSI), Durham,
NC. USA e-mail: wjliao@math.duke.edu. Wenjing Liao is grateful to support from NSF-CCF-0808847, NSF-DMS-0847388,
ONR-N000141210601and SAMSI under Grant NSF DMS-1127914. 
}
}

\maketitle

\begin{abstract}
This paper presents a performance analysis of the MUltiple SIgnal Classification (MUSIC) algorithm applied on $D$ dimensional {\em single-snapshot} spectral estimation while $s$ true frequencies are located on the {\em continuum} of a bounded domain. Inspired by the matrix pencil form, we construct a D-fold Hankel matrix from the measurements and exploit its Vandermonde decomposition in the noiseless case. 
MUSIC amounts to identifying a noise subspace, evaluating a noise-space correlation function, and  localizing frequencies by searching the $s$ smallest local minima of the noise-space correlation function.

In the noiseless case, $(2s)^D$ measurements guarantee an exact reconstruction by MUSIC as the noise-space correlation function vanishes exactly at true frequencies. When noise exists, we provide an explicit estimate on the perturbation of the noise-space correlation function in terms of noise level, dimension $D$, the minimum separation among frequencies, the maximum and minimum amplitudes while frequencies are separated by $2$ Rayleigh Length (RL) at each direction. 
As a by-product the maximum and minimum non-zero singular values of the multidimensional Vandermonde matrix whose nodes are on the unit sphere are estimated under a gap condition of the nodes. Under the 2-RL separation condition, if noise is i.i.d. gaussian, we show that perturbation of the noise-space correlation function decays like $\sqrt{\log(\#(\bN))/\#(\bN)}$ as the sample size $\#(\bN)$ increases.  

When the separation among frequencies drops below 2 RL, our numerical experiments show that the noise tolerance of MUSIC obeys a power law with the minimum separation of frequencies.  
\end{abstract}


%
\IEEEpeerreviewmaketitle

\section{Introduction}

\subsection{Problem formulation}

The problem of estimating the spectrum of a signal arises in many fields of application, such as array imaging \cite{KV96,SAS}, Direction-Of-Arrival (DOA) estimation \cite{SchD}, sonar and radar \cite{CB}, remote sensing \cite{FSY}, inverse scattering \cite{FannjiangII}, geophysics \cite{Bath}, etc. In many situations signals are inherently multidimensional which presents a challenging set of theoretical and computational difficulties that must be tackled \cite{MultiSE}.


\commentout{
A large class of applications in science and engineering features signals whose spectrum can be modeled by superposition several spikes in the continuum of a bounded domain, and involves estimation of the underlying spectrum from finite samples of the signal. Examples include array imaging \cite{KV96}, sonar and radar \cite{CB}, remote sensing \cite{FSY}, inverse scattering \cite{BPTB}, geophysics \cite{Bath}, etc. In many situations signals are inherently multidimensional, which presents a challenging set of theoretical and computational difficulties that must be tackled.
}

Specifically the multidimensional spectral estimation problem aims to recover the spectrum of a multidimensional signal from a collection of its time-domain samples. Let $\bt,\bom \in \RR^D$. Consider a signal $y(\bt)$ consisting of a linear combination of $s$ time-harmonic components from the set 
$$\{e^{2\pi i \bom^j\cdot \bt} : \bom^j  =(\om^j_1,\ldots,\om^j_D)\in \RR^D, \ j = 1,\ldots,s\},$$
i.e.,
$$ y(\bt) = \sum_{j=1}^s x_j e^{2\pi i \bom^j \cdot \bt}.$$
In the noisy model,
\beq
\label{model}
y^e(\bt) = y(\bt) + e(\bt), 
\eeq
where $e(\bt)$ represents noise. 

Our task is to find out the frequency support $\supp =	\{\bom^1, . . . , \bom^s\}$ 	and	the	corresponding	amplitudes	$\bx	=	[x_1, . . . , x_s]^T$ from $y^e(\bt)$ sampled at $\bt  =\bn= (n_1,\ldots,n_D), \ 0\le n_k\le N_k, k=1,\ldots,D$. Sampling in $\bt$-domain results in periodicity in $\bom$-domain so we can only hope to determine frequencies on the torus $\TT^D = [0,1)^D$ with wrapped distance. For $k=1,\ldots,D$, we define the wrapped distance between $\bom^j$ and $\bom^l$ at the $k$-th direction to be
\beq
\label{distance}
{\itd}_k(\bom^j,\bom^l) = \min_{m\in\ZZ}  | \om_k^j+m-\om_k^l |.
\eeq 
Due to the nonlinear dependence of $y(\bt)$ on $\supp$, the main difficulty  of spectral estimation lies in identifying the support set $\supp$. The amplitudes $\bx$ can be retrieved by solving a least-square problem once $\supp$ has been found. 

A key unit in classical resolution theory \cite{rs} is the Rayleigh Length (RL), which is considered as the minimum resolvable separation of two objects with equal intensities. One RL $=1/N$ in one dimension. In the multidimensional case, we consider anisotropic sampling with possibly distinct $N_k,\  k=1,\ldots,D$
and let $1/N_k$ be the RL at the $k$-th direction.


\subsection{Motivation}
After the emergence of compressive sensing \cite{CandesCS,DonohoCS}, spectral estimation with an emphasis on sparsity  was explored in various applications, such as medical imaging \cite{CandesCS}, remote sensing \cite{FSY} and inverse scattering \cite{FannjiangII}.
With few exceptions frequencies considered in the compressive sensing community are assumed to be on the DFT grid with grid spacing = 1 RL. However, spectrum of a natural signal is composed of a few frequencies lying on a continuum, and they can not be well approximated by atoms on the DFT grid. The gridding error \cite {DB,FL,FL1} or basis mismatch \cite{Chi} is significant on the DFT grid, manifesting the gap between the continuous world and the discrete world.


This motivates us to explore algorithms which are capable of recovering frequencies located on a continuum.
The MUltiple Signal Classification (MUSIC) algorithm proposed by Schmidt \cite{SchD,Sch} is a classical high-resolution DOA estimator, but its performance analysis for the off-the-grid frequencies is lacking in literature.
In one dimension the authors provided a stability analysis for MUSIC while frequencies are on a continuum with a minimum separation $\ge $ 2RL, and explored the super-resolution limit of MUSIC in \cite{FLMUSIC}. 
In many applications signals are inherently multidimensional which motivates us to study the performance of MUSIC in higher dimensions.


\subsection{Main results}
MUSIC was originally proposed in \cite{SchD,Sch} as a DOA estimator so it works on the Multiple Measurement Vector (MMV) problem in which the measurements contain multiple snapshots of distinct signals whose spectra share a common support. 
In this paper we consider spectral estimation with a single snapshot of measurement which also has a variety of important applications.
Inspired by the convectional matrix pencil form \cite{Rao1,MatrixPencil2,DHankel}, we form a D-fold Hankel matrix from the measurements $\{y^e(\bn), 0\le n_k \le N_k\}$ and exploit its Vandermonde decomposition in the noiseless case to rewrite the problem in the form of MMV. The MUSIC algorithm amounts to identifying a noise subspace, forming a noise-space correlation function and searching the $s$ smallest local minima of the noise-space correlation function.

In the noiseless case we show that $(2s)^D$ measurements suffice to guarantee that the noise-space correlation function vanishes exactly at $\supp$ and therefore MUSIC gives rise to an exact reconstruction.
In the presence of noise, the D-fold Hankel matrix constructed from the noisy data $y^e$ is perturbed from the D-fold Hankel matrix of $y$ by the D-fold Hankel matrix of noise $e$ so that the noise-space correlation function no longer vanishes at $\supp$. The main contribution of this paper is to provide an explicit estimate on the perturbation of the noise-space correlation function in terms of noise level, dimension $D$, the minimum separation of frequencies, the maximum and minimum amplitudes under the 2RL-separation condition. Furthermore, if noise is i.i.d. gaussian, we show that MUSIC gets improved as $\bN$ increases, both analytically and numerically.

MUSIC also features its super-resolution effect, i.e., the capability of resolving frequencies separated below 1RL. Our numerical experiments in Section \ref{secnum2} show that, in the super-resolution regime, the noise tolerance of MUSIC follows a power law with the minimum separation of frequencies as the separation decreases.

\subsection{Comparison and connection to other works}

Among existing works, \cite{MUSIC} and \cite{FLMUSIC} are most closely related to this paper. In \cite{MUSIC} MUSIC was studied in the context of inverse scattering with random and sparse measurements when sparse frequencies are on a DFT grid. Later in \cite{FLMUSIC} the grid requirement was removed and a separation condition was added for MUSIC applied on one dimensional single-snapshot spectral estimation.
The present paper extends the stability analysis in \cite{FLMUSIC} to higher dimensions with the same emphasis that frequencies are off any grid.


In literature Prony was the first to address the spectral estimation problem \cite{Prony}, but the original 
Prony's method is numerically unstable. The best-known classical spectrum estimator is Fourier transform followed by localization which is shown to be much more sensitive to dynamic range $\xmax/\xmin$ than MUSIC in Section \ref{secnum}.




As mentioned earlier, our goal is to study algorithms which are capable of handling gridding error and achieving super-resolution. For the former purpose, some progresses have been made in greedy algorithms \cite{DB,FL} and analogs of $L_1$ minimization \cite{Csr1,Csr2,Tang}, etc. 

The greedy algorithms in \cite{FL} are based on techniques of band exclusion and local optimization proposed according to the coherence pattern of the sensing matrix when the problem is discretized on a fine grid. An approximate support recovery within 1RL is guaranteed for frequencies separated by 3RL.


In \cite{Csr1,Csr2}, Cand\`es and Fernandez-Granda proposed Total Variation (TV) minimization and showed that, in one dimension, while the frequencies are separated by 4 RL, TV-min yields an $L_1$ reconstruction error linearly proportional to noise level with a magnification factor as large as $F^2$ where F is the super-resolution factor \cite{Csr1}.  In regard to the two dimensional problem with isotropic sampling, i.e. $N_1=N_2$, it was proved that TV-min yields an exact recovery on the condition that frequencies are separated by 4.76RL, but the stability of TV-min was not clear, and the theory   has not been generalized to dimension $\ge 3$.

For sparse frequencies, Tang et al. \cite{Tang} proposed atomic norm minimization for the one dimensional problem and showed that exact reconstruction is guaranteed under the support separation condition $\ge $ 4RL while $\mathcal{O}(s\log s\log N)$ measurements are taken at random. In \cite{Chi2D}, a similar performance guarantee for the atomic norm minimization was established in two dimensions. Additionally the amplitudes are assumed to have random phases and the noise sensitivity is not considered in \cite{Tang,Chi2D}.  In comparison, the uniqueness result for MUSIC in our Theorem \ref{thm3} requires $(2s)^D$ measurements but does not enforce any separation condition or random-phase condition for the amplitudes, and furthermore stability of MUSIC is established under the 2-RL separation condition.

As for numerical implementations of the atomic norm minimization, a SemiDefinite Programming (SDP) on the dual problem was solved in \cite{Tang,Csr2} in one dimension. In higher dimensions the SDP with noise-free measurements was approximately formulated in \cite{Chi2D} and later exactly formulated in \cite{Xu2D}. Unfortunately the current SDP in \cite{Xu2D} does not handle noise so we only test it in the noiseless case. In applications the computational inefficiency could be a concern for SDP. 

Thanks to anonymous referees, we became aware of Moitra's work on the performance analysis of the one dimensional matrix pencil method \cite{Moitra} in which the best-matching frequency error and amplitude error are estimated in terms of noise level and the minimum singular value of a Vandermonde matrix. 
The conditioning of the Vandemonde matrix was studied in \cite{FLMUSIC} through the discrete Ingham inequality \cite{Ingham} and in \cite{Moitra} through the discrete large sieve inequality \cite{Vaaler1985}, respectively. In one dimension the discrete large sieve inequality offers tighter bounds than the discrete Ingham inequality due to the use of extremal functions. Theorem \ref{thm2} in the current work can be viewed as a multidimensional discrete version of the large sieve inequality.


Many recent works in spectral estimation focus on super-resolution.
It was proved in \cite{MC} that, in one or two dimensions, if all amplitudes are {\em real positive} and the support set has Rayleigh index $r$ \cite{Donoho92,MC}, solving a convex feasibility problem yields an $L_1$ reconstruction error linearly proportional to noise with a magnification factor as large as $F^{2r}$ where $F$ is the super-resolution factor. While positivity constraint is crucial in \cite{MC}, the super-resolution of MUSIC is demonstrated for complex-valued amplitudes in Figure \ref{figsup3}.

As a leading candidate for DOA estimation, MUSIC has attracted many performance studies \cite[etc]{SN,Friedlander,LLV,Swindlehurst1,Swindlehurst2} in the past 30 years. In DOA estimation, infinite snapshots of measurements for signals whose spectra share a common support 
are taken and the problem is rewritten in the form of MMV by forming the covariance matrix of the measurements. The errors arising from approximating the covariance matrix with finite samples and a perturbed imaging vector were analyzed in \cite{SN} and \cite{Friedlander}, respectively, under the assumption that the covariance matrix of noise is a multiple of the identity matrix so that the noise subspace is not affected by noise. The implication of the results in \cite{SN} and \cite{Friedlander} on the single-snapshot spectral estimation problem \eqref{model} is not clear because the D-fold Hankel matrix of noise $e$ never appears as a multiple of identity. In \cite{LLV}, the constraint on the statistics of noise was removed and sensitivity of MUSIC was studied in terms of the minimum non-zero singular value of the covariance matrix constructed from the noiseless data, which is, however, an implicit quantity depending on the support set $\supp$ and the amplitudes $\bx$. In comparison, the perturbation estimate in the present work is explicitly given in terms of noise level, dimension $D$ and the minimum separation of frequencies when the separation is above $2$ RL.

\subsection{Organization of the paper}
The rest of paper is organized as follows. In Section \ref{secmusic}, we introduce the D-fold Hankel matrix and the MUSIC algorithm. Section \ref{secvan} is devoted to a study on the multidimensional Vandermonde matrix which plays a significant role in the application of MUSIC. In Section \ref{secper}, we present a performance guarantee of MUSIC, including a sufficient condition for exact reconstruction in the noiseless case and a stability analysis in the presence of noise. A systematic numerical experiment on the stability and super-resolution of MUSIC is provided in Section \ref{secnum}. Finally we conclude and discuss possible extensions of the current work in Section \ref{seccon}. Proofs of theorems are left in the Appendix.

\subsection{Notations} 

Here we define some notations to be used in subsequent sections. For a vector $\bx  = \{x_j\}_{j=1}^D \in \CC^{D}$, the $p$-norm of $\bz$ for $0 < p < \infty$ is $\|\bx\|_p = (\sum_{j=1}^D |x_j|^p)^{1/p}$. When $p =\infty$, $\|\bx\|_\infty = \max_{j=1}^D |x_j|$. 
Let $\xmax = \max_{j=1}^s |x_j|$ and $\xmin = \min_{j=1}^s |x_j|$. The dynamic range of $\bx$ is defined as $\xmax/\xmin$.
For an $m \times n$ matrix $A$, $A^T$ and $A^*$ denote the transpose and conjugate transpose of $A$.
$\smax(A)$, $\smin(A)$ and $\si_j(A)$ represents the maximum, the minimum and the $j$-th singular value of $A$, respectively. Spectral norm of $A$ is $\|A\|_2 = \smax(A)$ and Frobenius norm of $A$ is $\|A\|_F = (\sum_{k=1}^m \sum_{j=1}^n |A_{k j}|^2)^{1/2}$.

Let $\bn = (n_1,\ldots,n_D)^T$, $\bN = (N_1,\ldots,N_D)^T$ and $\bL = (L_1,\ldots,L_D)^T$. $\bn \le \bN$ if $n_k \le N_k$ for $k = 1,\ldots,D$. Denote $\#(\bL) = (L_1+1)\cdots (L_D+1)$ and $\#(\bN-\bL) = (N_1-L_1+1)\cdots (N_D-L_D+1)$.
For $\bom = (\om_1,\ldots,\om_D) \in \TT^D$, the imaging vector of size $\bL$ at $\bom$ is defined to be
\beq
\label{dphiL}
\phi^\bL(\bom) 
 =  \text{vec}(\{e^{2\pi i \bn\cdot \bom} , \mathbf{0} \le \bn \le \bL\})
 \in \CC^{\#(\bL)}
\eeq
where ${\rm vec}(\cdot)$ is the operation of converting an array into a column vector in the order such that
$$
\left[\text{vec}(\{e^{2\pi i \bn\cdot \bom} , 0\le \bn \le \bL\})\right]_{m+1} = e^{2\pi i( n_1\om_1 +\ldots + n_D \om_D)} ,
$$
where
$n_1 = \left\lfloor {m}/{\#(L_2,\ldots,L_D)} \right\rfloor,$
$ r_1 = m$ mod $\#(L_2,\ldots,L_D)$,
$ n_2 = \left\lfloor {r_1}/{\#(L_3,\ldots,L_D)}\right\rfloor,$
$r_2 = r_1$ mod $\#(L_3,\ldots,L_D)),$ $\ldots,$
and
$n_D = \left\lfloor {r_{D-1}}/{L_D+1}\right\rfloor,$ $r_D = r_{D-1} \text{ mod } L_D+1$ for $m =0 ,\ldots, (L_1+1)\cdots (L_D+1)-1$.

Suppose the exact frequency support is $\supp = \{\bom^j , j=1,\ldots,s\}\subset \TT^D$ where the superscript $j$ identifies the $j$-th frequency and the subscript $k$ refers to the $k$-th coordinate.
We
define the matrix
$
\Phi^{\bL} 
= 
\begin{bmatrix}
\phi^{\bL}(\bom^1) 
& 
\phi^{\bL}(\bom^2)
& 
\ldots
&
\phi^{\bL}(\bom^s)
\end{bmatrix}.
$
For $k = 1,\ldots,D$ we let $\Psi^{(k,L_k)}$ be the Vandermonde matrix with nodes $\{e^{2\pi i \om^j_k}, j=1,\ldots,s\}$, i.e.,
$$\Psi^{(k,L_k)} = 
\begin{bmatrix}
e^{2\pi i\om^1_k\cdot 0}  & e^{2\pi i\om^2_k \cdot 0} & \ldots & e^{2\pi i\om^s_k \cdot 0}\\
\commentout{
e^{2\pi i\om^1_k \cdot 1} & e^{2\pi i\om^2_k \cdot 1} & \ldots &e^{2\pi i\om^s_k \cdot 1}\\
}
\vdots  & \vdots & \vdots & \vdots\\
e^{2\pi i\om^1_k\cdot L_k} & e^{2\pi i\om^2_k\cdot L_k} & \ldots &e^{2\pi i\om^s_k\cdot L_k}
\end{bmatrix}.$$


\section{MUSIC for multidimensional spectral estimation}
\label{secmusic}

MUSIC works on the MMV problem.
When only one snapshot of measurement is available, we form a D-fold Hankel matrix \cite{DHankel} based on data $\{y^e(\bn), \ \mathbf{0} \le \bn \le \bN\}$ and exploit its Vandermonde decomposition in the noiseless case.

\subsection{D-fold Hankel matrix}
\label{sechankel}


We first consider the two dimensional case $D =2$.
\begin{align*}
& y_{n_1,n_2}  = y(n_1,n_2) = \sum_{j=1}^s x_j e^{2\pi i (n_1 \om^j_1 + n_2\om^j_2)}, \\
& \ 0\le n_1\le N_1 , 0 \le n_2 \le N_2.
\end{align*}
Fixing a positive integer $L_2 < N_2$, we form the Hankel matrix
$$
A_{n_1} = 
\begin{bmatrix}
y_{n_1,0} & y_{n_1,1} & \ldots &y_{n_1,N_2-L_2} \\
y_{n_1,1} & y_{n_1,2} & \ldots &y_{n_1,N_2-L_2+1}\\
\vdots & \vdots & \vdots & \vdots \\
y_{n_1,L_2} & y_{n_1,L_2+1} & \ldots & y_{n_1,N_2}
\end{bmatrix}.
$$
for every $0 \le n_1 \le N_1$.
Let
$
\Lambda_1  = \diag(e^{2\pi i \om^1_1}, \ldots, e^{2\pi i \om^s_1})$ and $
X  = \diag(x_1,x_2,\ldots,x_s)$.
Then $A_{n_1}$ can be decomposed as
$$
A_{n_1} = \Psi^{(2,L_2)} X \Lambda_1^{n_1}{\Psi^{(2,N_2-L_2)}}^T.
$$
Fixing another positive integer $L_1<N_1$, we build the Hankel block matrix
$$
H =\Hankel(y)
=
\begin{bmatrix}
A_0 & A_1 & \ldots & A_{N_1-L_1} \\
A_1 & A_2 & \ldots & A_{N_1-L_1+1} \\
\vdots & \vdots & \vdots & \vdots \\
A_{L_1} &A_{L_1+1} &   \ldots & A_{N_1} 
\end{bmatrix}.
$$
which can be decomposed as
{\small
$$
H = 
\begin{bmatrix}
\Psi^{(2,L_2)} \Lambda_1^0 \\
\vdots \\
\Psi^{(2,L_2)} \Lambda_1^{L_1}
\end{bmatrix}
X
\begin{bmatrix}
\Lambda_1^0 {\Psi^{(2,N_2-L_2)}}^T 
&
\ldots
& 
\Lambda_1^{N_1-L_1} {\Psi^{(2,N_2-L_2)}}^T 
\end{bmatrix}.
$$}
According to \eqref{dphiL},
the Vandermonde decomposition of $H$ is exactly
$$
H =\Hankel(y)= \Phi^{\bL} X (\Phi^{\bN-\bL})^T,
$$
where $\bL = (L_1,L_2)^T$.

The D-fold Hankel matrix is a simple extension of the 2-fold Hankel matrix with recursion. Consider
\begin{align*}
& y_{n_1,\ldots,n_D}  = y(n_1,\ldots,n_D) = \sum_{j=1}^s x_j e^{2\pi i (n_1 \om^j_1 + \ldots + n_D \om^j_D)}, \\
& \ 0\le n_k\le N_k ,  k=1,\ldots,D.
\end{align*}
Fixing integers $L_k< N_k$ for k = 1,\ldots,D, we define the D-fold Hankel matrix 
$$H = \Hankel(y) = 
\begin{bmatrix}
A_0 & A_1 & \ldots & A_{N_1-L_1} \\
A_1 & A_2 & \ldots & A_{N_1-L_1+1} \\
\vdots & \vdots & \vdots & \vdots \\
A_{L_1} &A_{L_1+1} &   \ldots & A_{N_1} 
\end{bmatrix}
$$
where $A_{l}$ denotes the (D-1)-fold Hankel matrix consisting of all entries in $\{y(l,n_2,\ldots,n_D), 0\le n_k \le N_k, k=2,\ldots,D\}$ with parameters $L_k, k=2,\ldots,D$. One can verify that $H$ has the following Vandermonde decomposition
\beq
\label{eqh6}
H = \Hankel(y) = \Phi^{\bL} X (\Phi^{\bN-\bL})^T
\eeq
with $\bL = (L_1,\ldots,L_D)^T.$

\subsection{The MUSIC algorithm}

After writing the problem in the form of MMV \eqref{eqh6}, we can apply the MUSIC algorithm to identify the support set. In the noiseless case, the crux of MUSIC is this: 

\begin{lemma}
\label{lem1}
If the following conditions are satisfied
\begin{enumerate}
\item $\rank(\Phi^{\bN-\bL}) = s$,
\item $\rank(\Phi^{\bL}) = s$,
\item $\rank([\Phi^{\bL} \ \phi^{\bL}(\bom)]) = s+1 $ for all $\bom \notin \supp$,
\end{enumerate}
Then $$\range(H) = \range(\Phi^{\bL}) = {\rm span}\{\phi^{\bL}(\bom^1),\ldots,\phi^{\bL}(\bom^s)\},$$ and the support set $\supp$ can be identified through the criterion:
$$\bom \in \supp \text{ if and only if } \phi^{\bL}(\bom) \in \range(H).$$
\end{lemma}

In practice MUSIC is realized through extracting peaks of the imaging function. If conditions in Lemma \ref{lem1} are satisfied, we compute the Singular Value Decomposition (SVD) of $H$ such that 
\begin{align*}
H = &
[\underbrace{U_1}_{\#(\bL) \times s} \ 
U_2] \
\underbrace{\text{diag}(\sigma_1,\ldots,\sigma_s,0,\ldots,0)}_{\#(\bL) \times \#(\bN-\bL)} 
\
 [\underbrace{V_1}_{\#(\bN-\bL) \times s} \ 
 V_2]^*
\end{align*}
with singular values $\sigma_1\ge \sigma_2\ge  \cdots\sigma_s>0.$ The column space of $U_1$ and $U_2$ are called signal space and noise space, respectively. Signal space is the same as $\range(H)$ and noise space is the null space of $H^*$. Let $\PO$ and $\PT$ be the orthogonal projections onto signal space and noise space respectively. For any $\bv \in \CC^{\#(\bL)}$,
$$\PO\bv = U_1(U_1^*\bv) \ \text{  and  } \
\PT\bv = U_2(U_2^*\bv).$$

Suppose conditions in Lemma \ref{lem1} are satisfied. Then $\bom \in \supp$ if and only if $\PT\phi^{\bL}(\bom) = \mathbf{0}$. Therefore the noise-space correlation function 
$$\CR(\bom) 
= \frac{\|\PT\phi^\bL(\bom)\|_2}{\|\phi^\bL(\bom)\|_2}
=\frac{\|U_2^* \phi^\bL(\bom)\|_2}{\|\phi^\bL(\bom)\|_2}$$
vanishes and the imaging function 
$$\CJ(\bom) = \frac{1}{\CR(\bom)}$$
peaks exactly on $\supp$.

In the presence of noise
$$H^e = \Hankel(\ye) = H + E = \Hankel(y) + \Hankel(e).$$
The problem becomes estimating the support set $\supp$ based on the noisy Hankel matrix $H^e$. Suppose the SVD of $H^e$ is
$$ H^e = [\underbrace{U^e_1}_{\#(\bL) \times s} \
U^e_2]
 \underbrace{\text{diag}(\se_1,\ldots,\se_s,\se_{s+1},\ldots)}_{\#(\bL) \times \#(\bN-\bL)} [\underbrace{V^e_1}_{\#(\bN-\bL) \times s} \ 
 V^e_2]^*$$
with singular values $\se_1\geq \se_2 \geq \cdots.$ 
The noise-space correlation function and the imaging function become 
$$
\RE(\bom) = \frac{\|\PET\phi^\bL(\bom)\|_2}{   \|\phi^\bL(\bom)\|_2} =  \frac{\|{U_2^e}^*\phi^\bL(\bom)\|_2}{\|\phi^\bL(\bom)\|_2} 
$$
and
$$ \JE(\bom) = \frac{1}{{\RE}(\bom)} 
$$
respectively with $\PET=U^e_2 {U^e_2}^*$.

The MUSIC algorithm is summarized in Table \ref{table1}.
\begin{table}
\begin{center}
   \begin{tabular}{|l|}\hline
    { \qquad\qquad\qquad\qquad\qquad\qquad {\bf MUSIC algorithm}} \\ \hline
    {\bf Input:} $\{y^e(\bn), \mathbf{0} \le \bn \le \bN\}, s, \bL$. \\
     1) Form the D-fold Hankel matrix $H^e = \Hankel(y^e) \in \CC^{\#(\bL) \times\#(\bN-\bL)  }$.
     \\
     2) SVD: $H^e = [U_1^e\  U_2^e] {\rm diag}(\se_1 , \ldots , \se_s ,\ldots) [V_1^e\ V_2^e]^* $, 
     \\
     \quad\ where $U_1^e \in \CC^{\#(\bL)\times s}$.\\
     3) Compute imaging function $\JE(\bom) = \|\phi^{\bL}(\bom)\|_2 /\|{U_2^e}^* \phi^\bL(\bom)\|_2$ 
     \\
     \quad\ for $\bom \in [0,1)^D$. \\
   {\bf Output:} $\widehat \supp =\{ \widehat\bom \text{ at the } s \text{ largest local maxima of } \JE(\bom) \} $.\\
    \hline
   \end{tabular}
\end{center}
\caption{MUSIC algorithm for multidimensional single-snapshot spectral estimation}
\label{table1}
\end{table}

\section{Multidimensional Vandermonde matrix}
\label{secvan}

Performance of MUSIC is crucially dependent on the rank and conditioning of the multidimensional Vandermonde matrix $\Phi^{\bL}$. To pave a way for the performance analysis of MUSIC, we discuss some properties of the multidimensional Vandermonde matrix $\Phi^\bL$ in this  section.

\subsection{Rank}
According to Lemma \ref{lem1}, we need to find conditions on the free parameters $\bL$ under which $\Phi^\bL$ has full column rank, i.e., $\rank(\Phi^\bL) = s$.
Suppose $\supp$ contains distinct frequencies.
The simplest case is in one dimension where the determinant of any square Vandermonde matrix is explicitly given. We have $\rank(\Phi^{L_1}) = s$ if and only if $L_1+1\ge s $. Unfortunately, due to the famous result of Mairhuber-Curtis \cite[Theorem 2.3]{Wendland}, we can not expect that $\rank(\Phi^{\bL}) = s$ when $(L_1+1)\cdots(L_D+1) \ge s$ in the multidimensional case $D>1$.
A sufficient condition to guarantee $\rank(\Phi^\bL) = s$ is 
$L_k +1 \ge s, \ \forall k=1,\ldots,D.$

\begin{lemma}
\label{lem2}
Suppose $\bom^j \neq \bom^l$ for all $\bom^j,\bom^l\in \supp, j\neq l$. Then $\rank(\Phi^\bL) = s$ if 
$$L_k +1 \ge s, \ \forall k=1,\ldots,D.$$
\end{lemma}

\begin{IEEEproof}
Since $\Psi^{(1,L_1)},\Psi^{(2,L_2)},\ldots,\Psi^{(D,L_D)}$ are submatrices of $\Phi^{\bL}$, we have 
\beq
\label{lem2eq}
\rank(\Phi^\bL) \ge 
\rank 
\begin{bmatrix}
\Psi^{(1,L_1)}
\\
\
\vdots
\\
\Psi^{(D,L_D)}
\end{bmatrix}.
\eeq
If $L_k + 1 \ge s, \forall k = 1,\ldots,D$, columns of of the matrix on the right hand side of \eqref{lem2eq} are linearly independent and therefore $\rank(\Phi^\bL) = s$.

\end{IEEEproof}

\subsection{Singular values}
The conditioning of the multidimensional Vandermonde matrix $\Phi^\bL$ plays a significant role in the stability of MUSIC to be discussed in Section \ref{secper}. 
Motivated by the large sieve inequality \cite{Vaaler1985},
we provide an estimate on the singular values of $\Phi^\bL$ in the case that the support set $\supp$ satisfies a gap condition in Theorem \ref{thm2}. In one dimension, Theorem \ref{thm2} improves the discrete Ingham inequality in \cite{FLMUSIC} in terms of tightness due to the use of extremal functions.


\begin{theorem}
\label{thm2}
 Let $L_1,\ldots,L_D$ be even integers. If $\supp = \{\bom^1,\ldots,\bom^s\}\subset \TT^D$ fulfills the gap condition 
\beq
\itd_k(\bom^{j},\bom^l) \ge q_k > \frac{1}{L_k}, \ \forall j \neq l,\ k=1,\ldots,D,
\label{sepding}
\eeq
then 
\begin{align}
&\|\bc\|^2 
\prod_{k=1}^D \left( 1-\frac{1}{q_k L_k}
\right)
\le
\frac{\|\Phi^\bL \bc\|^2}{\#(\bL-\mathbf{1})}
\le
\|\bc\|^2 
\prod_{k=1}^D \left( 1+\frac{1}{q_k L_k }
\right)
 \label{ding}
 \end{align}
 for all square-summable $\bc = \{c_j\}_{j=1}^s \in \CC^s$.
 In other words,
\begin{align}
\frac{\smax^2(\Phi^\bL) }{\#(\bL-\mathbf 1)} 
&\le  \prod_{k=1}^D \left( 1+\frac{1}{q_k L_k}
\right),
\label{smax}
\\
\frac{\smin^2(\Phi^\bL)}{\#(\bL-\mathbf 1)} 
& \ge \prod_{k=1}^D \left( 1-\frac{1}{q_k L_k}
\right).
\label{smin}
\end{align}
\end{theorem}

The proof of Theorem \ref{thm2} is in Appendix \ref{app1}.

\commentout{
\begin{remark}
There are three main differences between Theorem \ref{thm2} and Proposition \ref{propingham}. First, Proposition \ref{propingham} and Theorem \ref{thm2}, respectively, consider the continuous complex exponentials $\{e^{2\pi i \bom^j \cdot \bt}\}$ for $\bt \in B^2_{L/2}$ and the discrete complex exponentials $\{e^{2\pi i \bom^j \cdot \bn}\}$ for $\bn \in \calL^D(L)$. Accordingly, the integral in \eqref{ing} is replaced by a discrete sum in \eqref{ding}. Second, in Theorem \ref{thm2}, $\bn$ is taken on the lattice $\calL^D(L)$ whose continuous analogy is $B^\infty_{L/2}$ (up to a time shift) while $\bt$ in Proposition \ref{propingham} is over $B^2_{L/2}$. As a result, the gap conditions \eqref{seping} and \eqref{sepding} differ by a factor of $\sqrt{D}$. Third, discrete multidimensional Ingham inequalities in Theorem \ref{thm2} not only prove the existence of the upper and lower bounds but also explicitly deliver the constants in terms of $L,q$ and $D$.
\end{remark}
}

\begin{remark}
The upper bound in \eqref{ding} holds even when the gap condition \eqref{sepding} is violated; however, \eqref{sepding} is necessary for the positivity of the lower bound.
\end{remark}

\begin{remark}
\label{remark2}
Theorem \ref{thm2} is stated for even integers $L_1,\ldots,L_D$, but a slightly looser bound also holds if some of them are odd. Without loss of generality, we assume $L_1,\ldots,L_m$ are odd and $L_{m+1},\ldots,L_D$ are even for some integer $m$. Let $\bL^+ = (L_1+1,\ldots,L_m+1,L_{m+1},\ldots,L_D)^T$ and $\bL^- = (L_1-1,\ldots,L_m-1,L_{m+1},\ldots,L_D)^T$.
One can expect $\smax^2(\Phi^\bL)  \le \smax^2(\Phi^{\bL^+} )$ and $\smin^2(\Phi^\bL)  \ge \smin^2(\Phi^{\bL^-} )$.

\end{remark}

\section{Performance guarantee of MUSIC}
\label{secper}

In this section we provide a performance guarantee for the  MUSIC algorithm, including a sufficient condition for exact reconstruction in the noiseless case and a perturbation estimate on the noise-space correlation function in the presence of noise.

In the noise-free case, combining Lemma \ref{lem1} and Lemma \ref{lem2} yields a sufficient condition for exact recovery by MUSIC.

\begin{theorem}
\label{thm3}
Suppose $\bom^j \neq \bom^l$ for all $\bom^j,\bom^l\in \supp$ with $j\neq l$. If
\beq
\label{thm3e1}
L_k \ge s \text{ and } N_k -L_k+1 \ge s, \ \forall k = 1,\ldots,D,
\eeq
then
$$\bom \in \supp \Longleftrightarrow \CR(\bom) = 0 
\Longleftrightarrow
\CJ(\bom) = \infty.$$
MUSIC algorithm yields an exact reconstruction of $\supp$.
\end{theorem}

\begin{remark}
Condition \eqref{thm3e1} gives a sufficient condition for the success of MUSIC, saying that the number of measurements $\prod_{k=1}^D(N_k+1) \ge (2s)^D$ suffices to guarantee an exact reconstruction. 
\end{remark}

In the presence of noise the noise-space correlation function $\RE(\bom)$ no longer vanishes at $\supp$ and the imaging function $\JE(\bom)$ no longer peaks at $\supp$. Performance of MUSIC relies on how much the noise-space correlation function is perturbed from $\CR(\bom)$ to $\RE(\bom)$. 
 In \cite{FLMUSIC}  a perturbation estimate of the noise-space correlation function was given in terms of $\si_1(H)$, $\si_s(H)$ and $\|E\|_2$.

\begin{lemma}
\label{lemmap1}
Suppose $\si_s(H) > 0$ and $\|E\|_2 < \si_s(H)$. Then
\beq
\label{eqp}
|\RE(\bom) -\CR(\bom)| \le
 \frac{4\si_1(H)+2\|E\|_2}{(\si_s(H) -\|E\|_2)^2} \|E\|_2.
\eeq
\end{lemma}

Theorem \ref{lemmap1} holds for  all signal models with any support set $\supp$ as long as $\si_s(H) > 0$ and $\|E\|_2 < \sigma_s(H)$. In order to make the bounds \eqref{eqp} meaningful, we apply Thoerem \ref{thm2} to derive an explicit  perturbation estimate while $\supp$ satisfies a gap condition.

\begin{theorem}
\label{thm4}
Let $\bN = (N_1,\ldots,N_D)^T$ and  $\bL = (L_1,\ldots,L_D)^T$. Suppose $L_k$ and $N_k-L_k, \ k=1,\ldots,D$, are even integers. 
If the support set $\supp = \{\bom^1,\ldots,\bom^s\}$ satisfies the gap condition 
\beq
\label{thm4e1}
q_k=\min_{j\neq l}\itd_k(\bom_j,\bom_l)
 >\max\left(
 \frac{1}{L_k},
 \frac{1}{N_k-L_k}
\right),
  \eeq
then
{\small
\beq
\label{thm4e2}
|\RE(\bom)-\CR(\bom)| 
\le
\frac{4\alpha_1 + 2\frac{\|E\|_2}{\sqrt{\#(\bL)\#(\bN-\bL)}}}
{\left( \alpha_2 -\frac{\|E\|_2}{\sqrt{\#(\bL)\#(\bN-\bL)}} \right)^2}
\cdot
\frac{\|E\|_2}{\sqrt{\#(\bL)\#(\bN-\bL)}}
\eeq
}
where 
{\scriptsize
$$\alpha_1 =  \xmax \sqrt{\prod_{k=1}^D \frac{L_k(N_k-L_k)}{(L_k+1)(N_k-L_k+1)} \left(1+\frac{1}{q_k L_k} \right)\left(1+\frac{1}{q_k(N_k-L_k)} \right)},$$
$$\alpha_2 =  \xmin \sqrt{\prod_{k=1}^D \frac{L_k(N_k-L_k)}{(L_k+1)(N_k-L_k+1)}\left(1-\frac{1}{q_k L_k} \right)\left(1-\frac{1}{q_k(N_k-L_k)} \right)}.$$
}
\end{theorem}

\begin{remark}
Theorem \ref{thm4} is stated in the case that both $N_k$ and $L_k,\ k=1,\ldots,D$ are even integers. The results hold in other cases with a slightly different $\alpha_1$ and $\alpha_2$ based on Remark \ref{remark2}.
\end{remark}

\begin{remark}
In order to optimize the minimum separation in \eqref{thm4e1} we need to set $L_k = N_k/2$, which results in a square Hankel matrix. In this case the minimum separation in the $k$-th coordinate is $q_k \approx 2/N_k = 2$ RL, which implies that the resolving power of MUSIC in multidimensional case is at least $2$ RL.
\end{remark}

Proof  Theorem \ref{thm4} is in Appendix \ref{app3}. The key of Theorem \ref{thm4} is that, under the gap condition \eqref{thm4e1}, perturbation of the noise-space correlation function is proportional to $\|E\|_2/\sqrt{\#(\bL)\#(\bN-\bL)}$ with a magnifying constant about $4\alpha_1/\alpha_2^2$. 
Roughly speaking a large separation among frequencies at each direction and a close-to-unity dynamic range $\xmax/\xmin$ make a small magnifying constant and are therefore conducive to the stability of MUSIC.

The estimate \eqref{thm4e2} holds without any assumption on the statistics of noise. If $\{e(\bn)\}$ contains independent gaussian noise, i.e., $e \sim \calN(0,\sigma^2 I)$, we can further derive the asymptotics of $|\RE(\bom)-\CR(\bom)|$ with respect to $\bN$ and $\sigma$ as $\bN \to \infty$, i.e., $N_k \to \infty, \ k=1,\ldots,D$, based on the following estimate on $\|E\|_2$.

\begin{theorem}
\label{thm5}
Suppose the array $\{e(\bn), \ \mathbf{0} \le \bn \le \bN\}$ contains independent normal variables with mean $0$ and variance $\sigma^2$.
Then 
{\small
\beq
\label{thm5e1}
\EE \|E\|_2 \le  \sigma\sqrt{2 \max\{ \#(\bL),\#(\bN-\bL)\} \log(\#(\bL)+\#(\bN-\bL))}
\eeq
}
and furthermore, for all $t \ge 0$,
{\small
\begin{align*}
&
\PP\{\|E\|_2 \ge t\} 
\\
& \le (\#(\bL)+\#(\bN-\bL))\exp
\left(-\frac{t^2}{2\sigma^2 \max\{\#(\bL),\#(\bN-\bL)\}}\right).
\end{align*}
}
\end{theorem}


The proof of Theorem \ref{thm5} is in Appendix \ref{app2}.

Suppose frequencies in the support set are well separated at each direction such that $q_k = \min_{j \neq l} \itd_k (\bom_j,\bom_l) > 2/N_k, \ k=1,\ldots,D$. We set $\bL = \bN/2$ in the construction of Hankel matrix.
When noise is i.i.d. gaussian, $\EE\|E\|_2$ grows like $\mathcal{O}(\sigma\sqrt{\#(\bN)\log(\#(\bN))})$ as $\bN \to \infty$ which is  much smaller
than $\sqrt{\#(\bL)\#(\bN-\bL)} = \mathcal{O}(\#(\bN))$. As a consequence, $\EE\|E\|_2/\sqrt{\#(\bL)\#(\bN-\bL)} \ll \alpha_2 $ so the estimate \eqref{thm4e2} always holds, and moreover, 
\beq
\EE|\RE(\bom)-\CR(\bom)| =\mathcal{O}\left(
\sigma\sqrt{\frac{\log(\#(\bN))}{\#(\bN)}}
\right),
\label{asympN}
\eeq
which is numerically verified in Figure \ref{figN}.

\section{Numerical examples}
\label{secnum}

MUSIC is applied on the two-dimensional spectral estimation problem \eqref{model} with $N_1 = N_2 = N$ and i.i.d. Gaussian noise, i.e. $e \sim \mathcal{N}(0,\sigma^2 I) + i \mathcal{N}(0,\sigma^2 I)$. Since sampling is isotropic, RL $=1/N$ at each direction. Noise level is measured by 
\begin{center}
Noise-to-Signal Ratio (NSR) = $\frac{\EE \|e\|_F}{\|y\|_F} 
= \frac{\sigma \sqrt{2(N_1+1)(N_2+1)}}{\|y\|_F}
.$
\end{center}
Suppose sparsity $s$ is given. We test and compare the following methods:
\begin{enumerate}
\item MUSIC with $L_1= L_2 = L = N/2$. 
\item Fourier transform followed by localization: The Fourier transform of $\ye$ evaluated at $\bom \in [0,1)^2$ is
\beq
\label{inverseye}
\hbx(\bom) = \sum_{0 \le \bn \le \bN} \ye(\bn)e^{-2\pi i \bom\cdot \bn}.
\eeq

\item Atomic norm-min: The two dimensional atomic norm minimization is solved through the SemiDefinite Programming (SDP) formulated in \cite{Xu2D} with the application of CVX \cite{cvx}. It returns the dual polynomial $q(\bom)$ for $\bom \in [0,1)^2$. The current SDP in \cite{Xu2D} only considers the noise-free measurements, so we only test it in the noiseless case. 
\end{enumerate}

Recovered frequencies are extracted from the $s$ largest local maxima of the imaging function $\JE(\bom)$, $|\hbx(\bom)|$ and $|q(\bom)|$ respectively for the three algorithms above. 
Reconstruction error is measured by the Hausdorff distance between the exact support $\supp$ and the recovered support $\widehat\supp$:
$$\itd(\widehat\supp,\supp) 
=\max\left\{
\max_{\widehat{\bom} \in \widehat\supp} \min_{\bom\in\supp}\itd(\widehat\bom,\bom),
\max_{\bom \in \supp}
\min_{\widehat\bom \in \widehat\supp}
\itd(\widehat\bom,\bom),
\right\}$$
where $\itd(\widehat{\bom},\bom) = \max_{k=1}^D \itd_k(\widehat{\bom},\bom)$.

\subsection{Well-separated frequencies}
\label{secnum1}
\begin{figure*}[t]
\hspace{0.5cm}  MUSIC  \hspace{2.8cm}  Fourier transform + localization  \hspace{2cm} Atomic norm min
\centering
            \subfigure[Imaging function in log scale: $\log_{10}\JE(\bom)$]{
     \includegraphics[width=6cm]{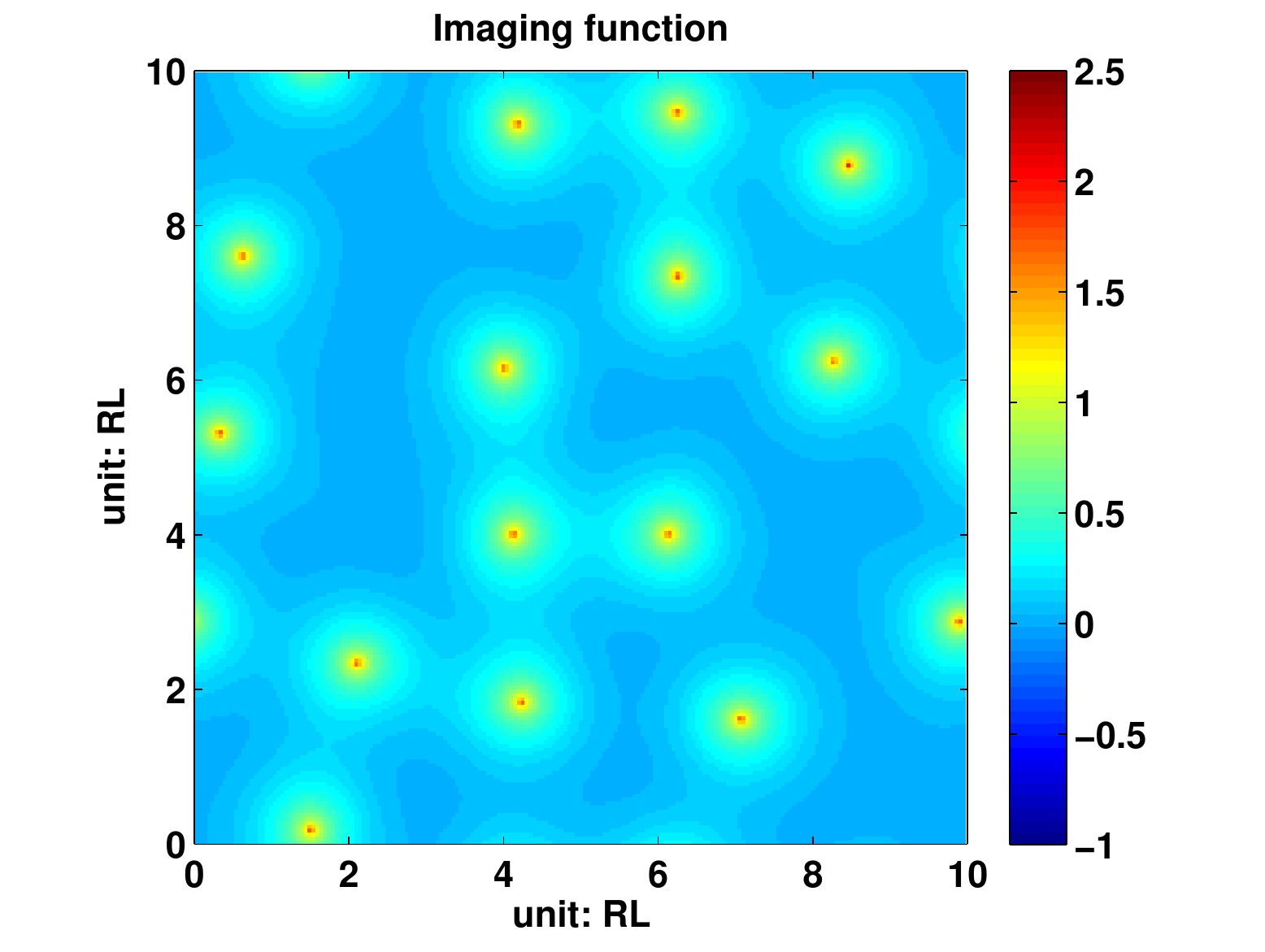}
     }
     \hspace{-0.6cm}
             \subfigure[$|\hbx(\bom)|$ in $\log_{10}$ scale: $\log_{10}|\hbx(\bom)|$]{
     \includegraphics[width=6cm]{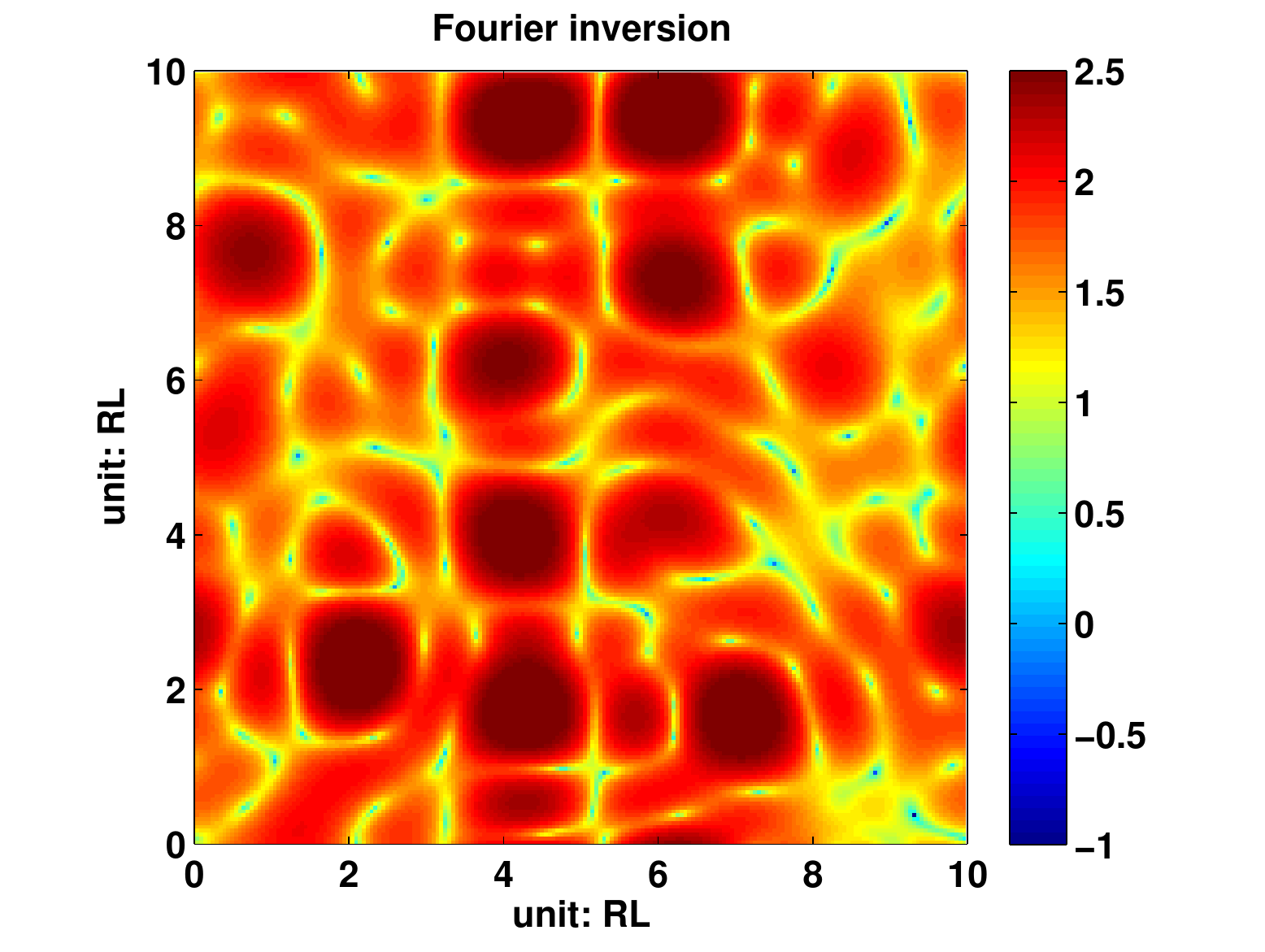}
     }
     \hspace{-0.6cm}
       \subfigure[Magnitudes of the dual polynomial: $|q(\bom)|$]{
     \includegraphics[width=6cm]{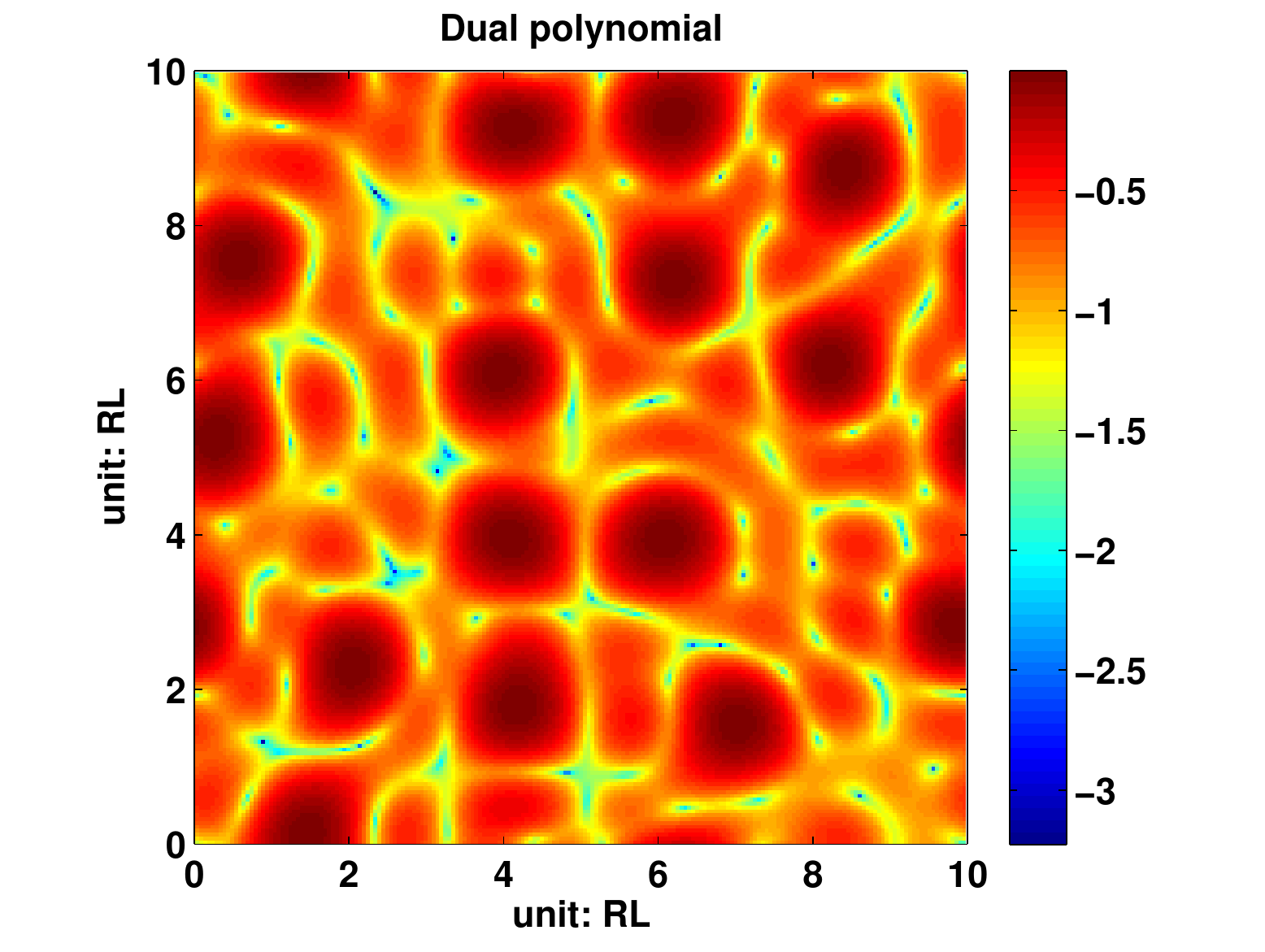}
     }
   \subfigure[Support recovery by MUSIC]{
     \includegraphics[width=6cm]{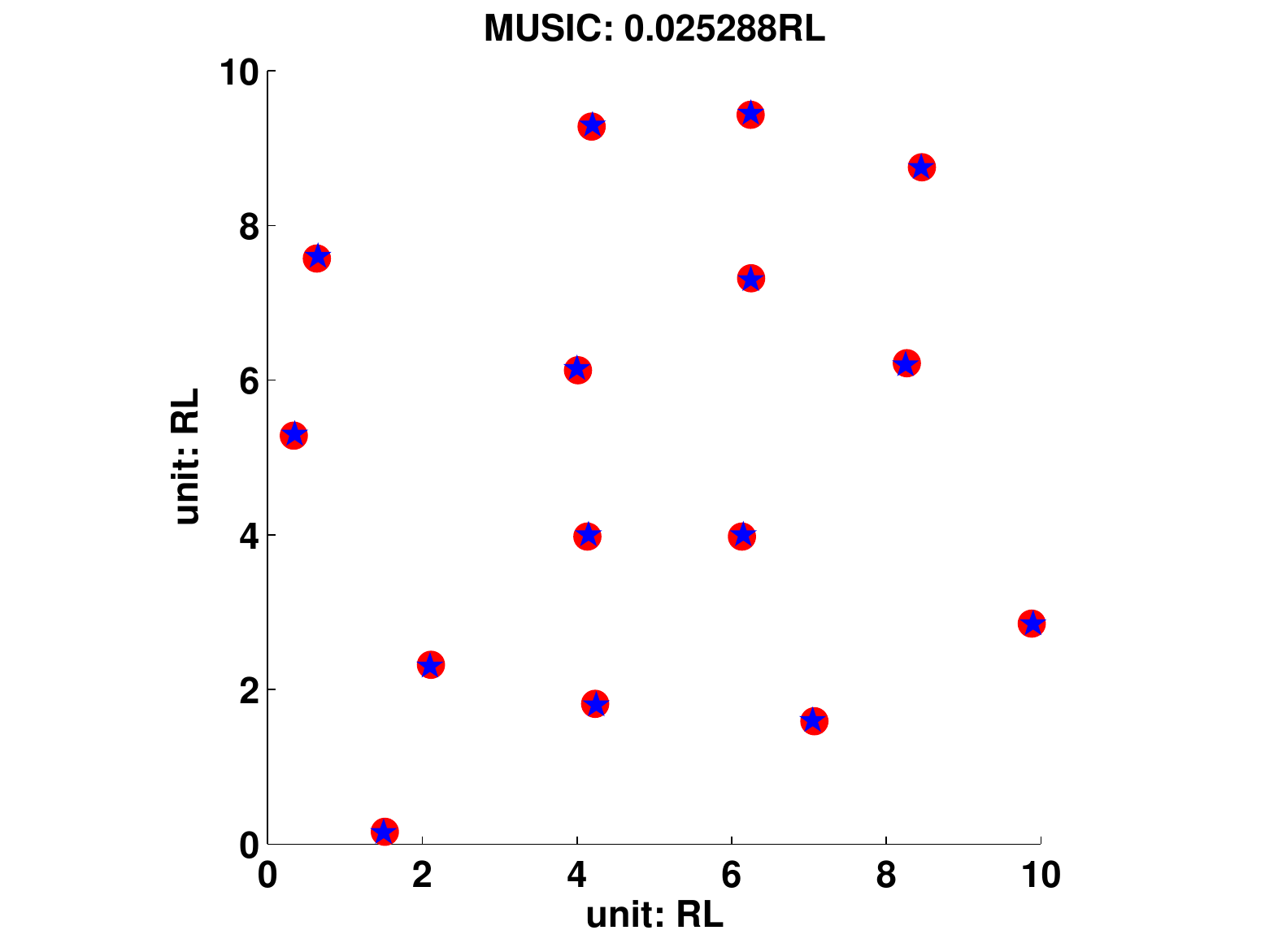}
     }
     \hspace{-0.6cm}
             \subfigure[Support recovery by Fourier transform + localization]{
     \includegraphics[width=6cm]{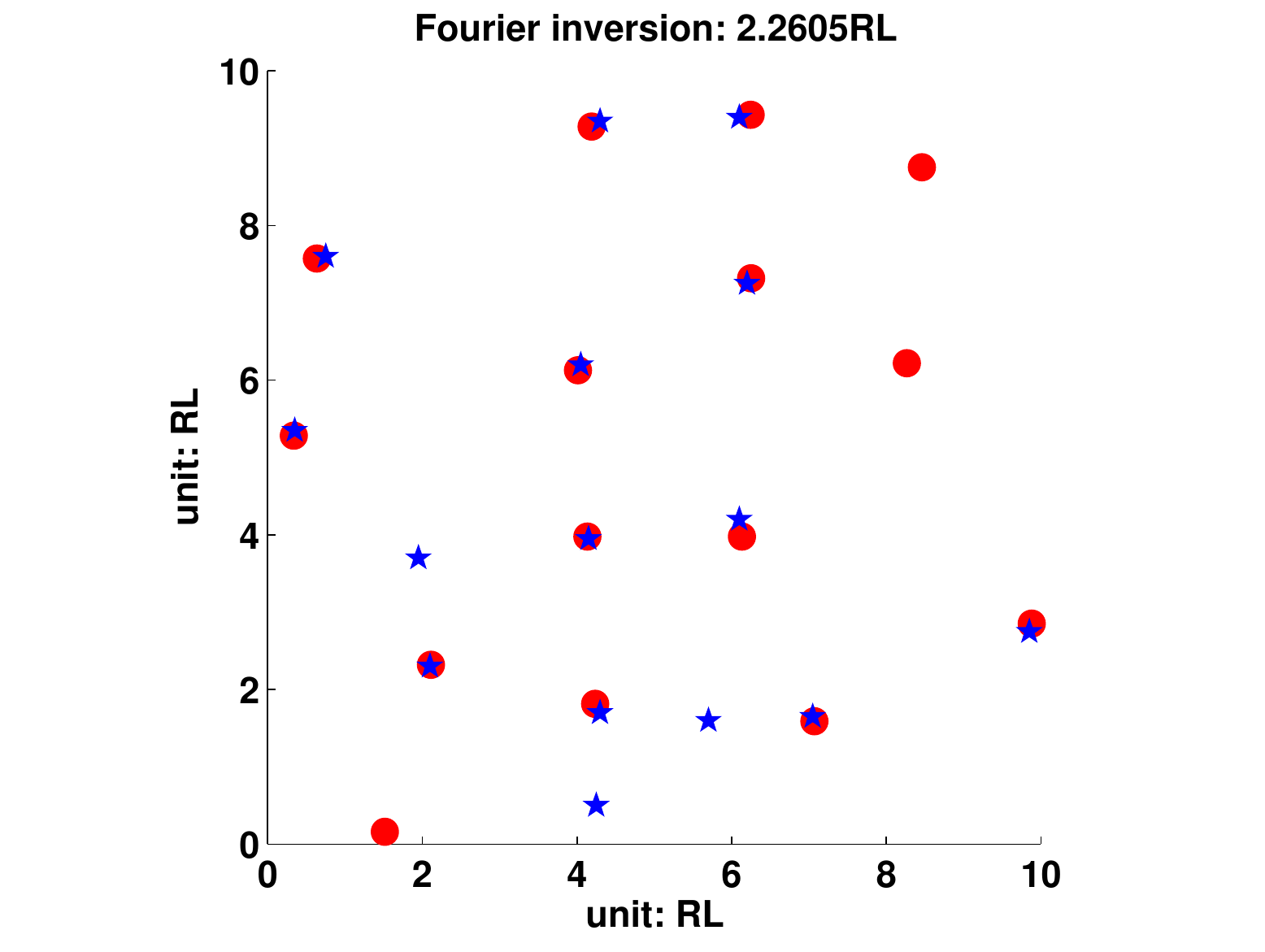}
     }
     \hspace{-0.6cm}
       \subfigure[Support recovery by atomic norm min]{
     \includegraphics[width=6cm]{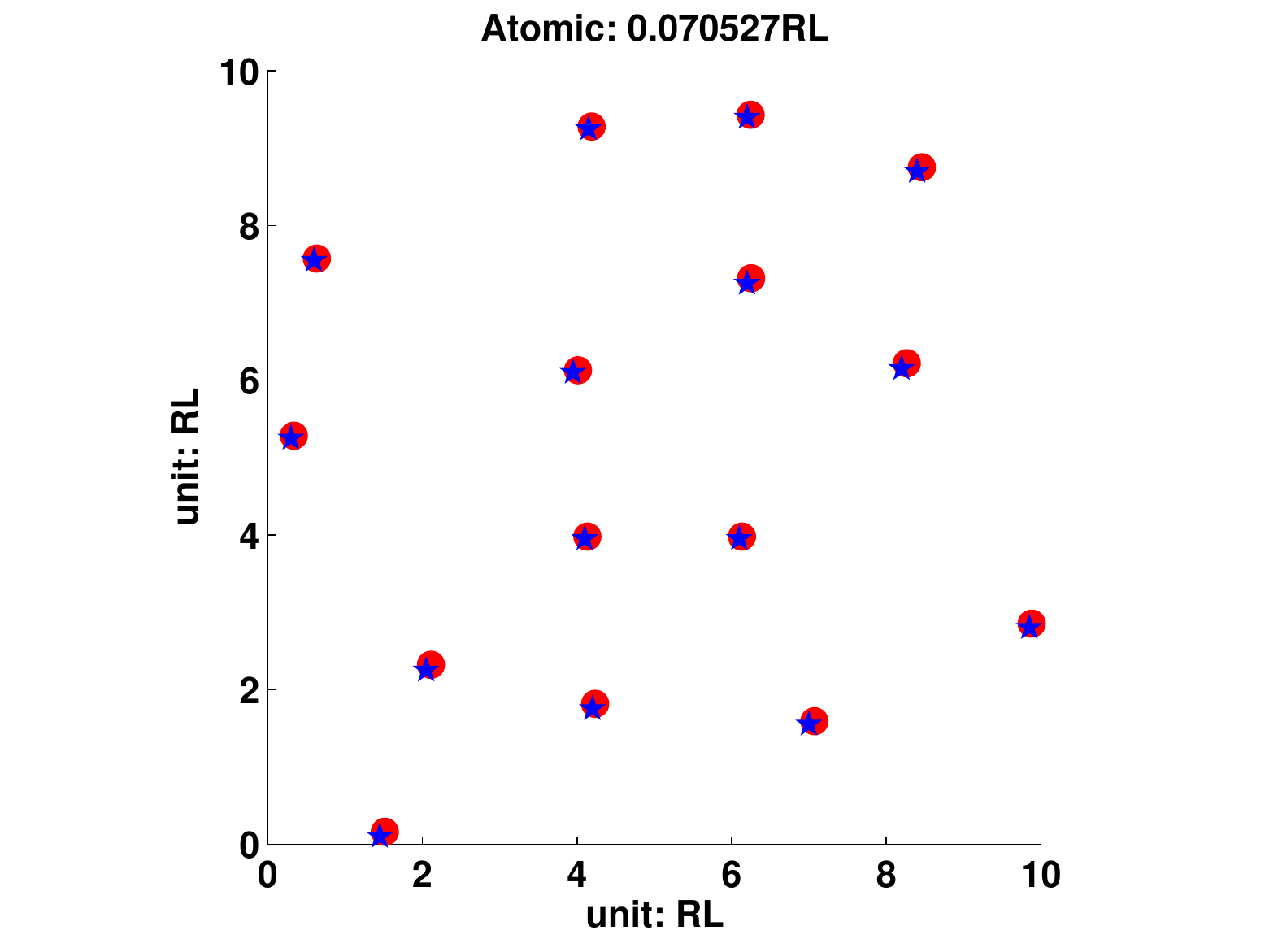}
     }
      \caption{In this example, $N = 10$, dynamic range = 5 and NSR = 0. 
      Three functions $\log_{10}\JE(\bom)$, $\log_{10}|\hbx(\bom)|$ and $|q(\bom)|$ are shown in (a)(b)(c) and support recoveries are shown in (d)(e)(f). In terms of support recovery, MUSIC and the atomic norm minimization achieve the accuracy of about $0.03$ RL and $0.07$ RL respectively, but Fourier transform followed by localization fails to detect three frequencies due to the slow decay of the Dirichlet kernel. In terms of running time, MUSIC and Fourier transform followed by localization take about 0.6396s and 0.7379s respectively but SDP  takes 49.3758s.
}
     \label{fig1}
\end{figure*}

First we compare the algorithms on the reconstruction of $15$ frequencies separated by 2RL or above when $N = 10$, NSR = 0 and $\bx$ has dynamic range $5$ and random phases. Results are displayed in Figure \ref{fig1}.
We observe that the imaging function $\JE(\bom)$ is more localized than $|\hbx(\bom)|$ in Fourier transform and the dual polynomial $|q(\bom)|$ in the atomic norm minimization. In terms of support recovery, MUSIC and the atomic norm minimization achieve the accuracy of about $0.03$ RL and $0.07$ RL respectively, but Fourier transform followed by localization fails to detect three frequencies. In terms of running time, MUSIC and Fourier transform followed by localization take about 0.6396s and 0.7379s respectively but SDP  takes 49.3758s, which implies that MUSIC is much more efficient than the atomic norm minimization.

\begin{figure}[hthp]
\centering
     \includegraphics[width=8cm,height=6cm]{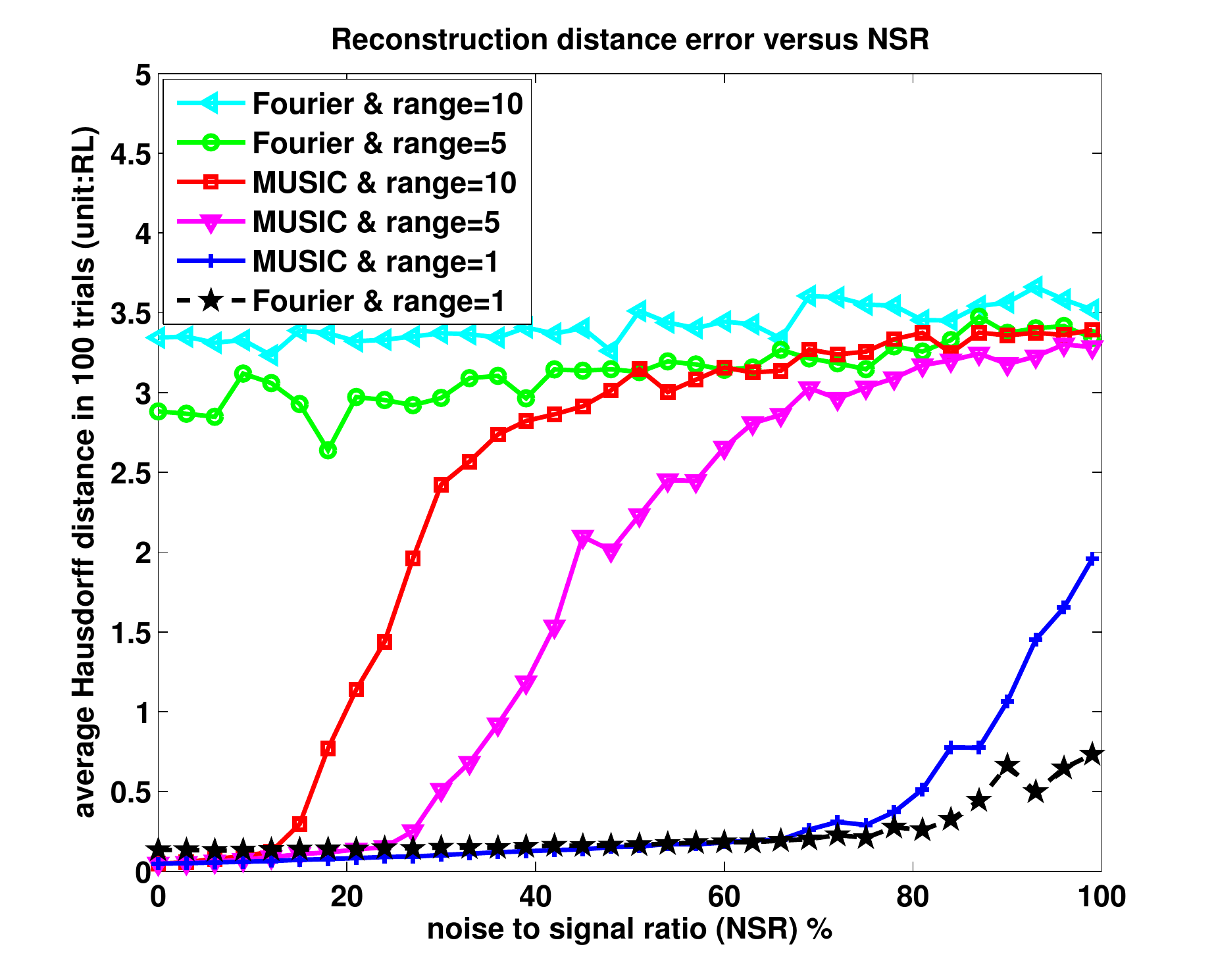}
     \caption{The average reconstruction error by MUSIC and Fourier transform followed by localization in 100 trials for $20$ frequencies separated by 2 RL with respect to varied NSR when the dynamic range of $\bx$ is 1, 5 and 10 respectively.
}
     \label{fig2}
\end{figure}

Then we compare the stability of MUSIC and Fourier transform followed by localization. $N=10$ is fixed. We adjust NSR by varying $\sigma$.
Figure \ref{fig2} shows the average error in 100 trials for the reconstruction of $20$ frequencies separated by 2 RL or above with respect to varied NSR in the case that $\bx$ has random phases and dynamic range 1, 5 and 10 respectively. 
The performance of both MUSIC and Fourier transform degrade as dynamic range increases. When dynamic range $=1$, both methods are very stable to noise, and Fourier transform followed by localization is slightly better when NSR is large. However, when dynamic range increases to $5$ and $10$, Fourier transform followed by localization fails even when NSR = 0, but MUSIC can handle $25\%$ and $10\%$ noise when dynamic range is $5$ and $10$ respectively.

\begin{figure}[hthp]
\centering
     \includegraphics[width=8cm,height=6cm]{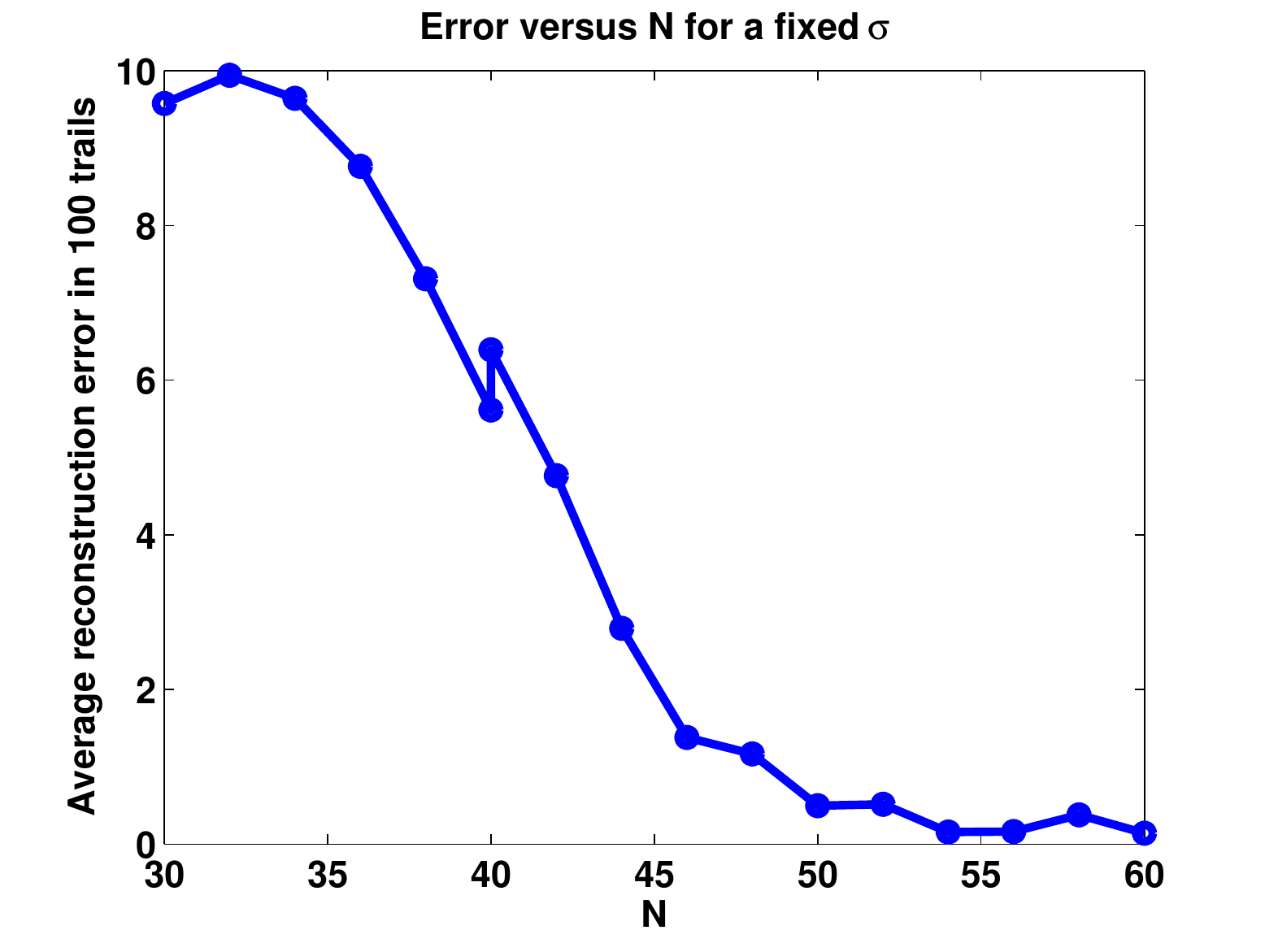}
     \caption{The average reconstruction error by MUSIC in 100 trials versus varied $N$ for $9$ frequencies located on a $3\times 3$ lattice whose side-length is 2RL. Noise $e \sim \mathcal{N}(0,\sigma^2 I) + i \mathcal{N}(0,\sigma^2 I)$ with $\sigma = 5$. As $N$ increases, NSR remains almost at the same level but the reconstruction error decreases, which is consistent with our analysis in \eqref{asympN}.
          }
     \label{figN}
\end{figure}

Our analysis in \eqref{asympN} demonstrates that, if noise $e$ contains i.i.d. gaussian random variables with fixed variance $\sigma^2$, perturbation of the noise-space correlation function decays like $\mathcal{O}(\sigma\sqrt{\log(\#(\bN))/\#(\bN)})$ as $\bN$ increases, which implies that the performance of MUSIC gets improved as $\bN$ becomes large. For a numerical verification, we fix a support set containing $9$ frequencies on a $3\times 3$ lattice whose side-length is 2RL and the amplitudes $\bx$ with dynamic range $1$. Noise $e \sim \mathcal{N}(0,\sigma^2 I) + i \mathcal{N}(0,\sigma^2 I)$ with $\sigma = 5$. In Fig. \ref{figN}, as $N$ increases, NSR remains almost at the same level (around 230\%) but the reconstruction error decreases, which is consistent with our analysis in \eqref{asympN}.

\subsection{Super-resolution of MUSIC}
\label{secnum2}
Super-resolution refers the capability of localizing frequencies separated below 1RL \cite{rs}. It has been numerically demonstrated in various applications \cite{KV96,SAS,Devaney,MUSIC,FLMUSIC} that MUSIC has super-resolution effect, but a mathematical theory is lacking. According to Lemma \ref{lemmap1}, super-resolution is possible if $\|E\|_2 \ll \sigma_s(H)$. As $\sigma_s(H) \ge \xmin \sigma_s(\Phi^\bL)\sigma_s(\Phi^{\bN-\bL})$, it is plausible that the noise tolerance of MUSIC obeys $\|E\|_2 \ll \xmin \sigma_s(\Phi^\bL)\sigma_s(\Phi^{\bN-\bL})$.

In one dimension, theory in \cite{Donoho92,DN15} suggests that $\si_s^2(\Phi^L)/L$ decreases to zero as $q$ (the minimum separation among frequencies) decreases to zero in a power law, and therefore, the noise tolerance of MUSIC follows a power law with respect to $q$, which has been numerically verified in \cite{FLMUSIC}.

In the multidimensional case, there is no theoretical result to explain the asymptotics of $\si_s^2(\Phi^\bL)/\#(\bL)$. Here we perform some numerical experiments to explore the relation between $\si_s^2(\Phi^\bL)/\#(\bL)$ and $q$ (the minimum separation among frequencies), and the relation between the noise tolerance of MUSIC and $q$, respectively.

\begin{figure}
\centering
\subfigure[Support A: $\supp_A$]{
\begin{tikzpicture}[xscale = 0.8,yscale = 0.8]
      \draw [-] (2,0) -- (0,0) ;
      \draw [-] (0,2) -- (0,0) ;
      \draw [fill] (2,0) circle [radius=0.1];
     \draw [fill] (0,0) circle [radius=0.1];
      \draw [fill] (0,1) circle [radius=0.1];
      \draw [fill] (1,0) circle [radius=0.1];
      \draw [fill] (0,2) circle [radius=0.1];
      \draw [white] (0,-0.2) circle [radius=0.1];
      \node at (1.5,0.2) {$q$};
      \node at (0.5,0.2) {$q$};
      \node at (-0.2,0.5) {$q$};
      \node at (-0.2,1.5) {$q$};
\end{tikzpicture}
}
\subfigure[Support B: $\supp_B$]{
\begin{tikzpicture}[xscale = 0.8,yscale = 0.8]
      \draw [-] (0,1) -- (0,0) ;
      \draw [-] (1,0) -- (0,0) ;
      \draw [-] (0,-1) -- (0,0) ;
      \draw [-] (-1,0) -- (0,0) ;
      \draw [fill] (0,1) circle [radius=0.1];
     \draw [fill] (0,0) circle [radius=0.1];
      \draw [fill] (0,-1) circle [radius=0.1];
      \draw [fill] (1,0) circle [radius=0.1];
      \draw [fill] (-1,0) circle [radius=0.1];
      \node at (0.5,-0.2) {$q$};
      \node at (-0.5,0.2) {$q$};
      \node at (0.2,0.5) {$q$};
      \node at (-0.2,-0.5) {$q$};
      \draw [white] (0,-1.2) circle [radius=0.1];
      \draw [white] (-1.2,0) circle [radius=0.1];
      \draw [white] (1.2,0) circle [radius=0.1];
\end{tikzpicture}
}
\subfigure[Support C: $\supp_C$]{
\begin{tikzpicture}[xscale = 0.8,yscale = 0.8]
      \draw [-] (2,0) -- (0,0) ;
      \draw [-] (-2,0) -- (0,0) ;
      \draw [fill] (2,0) circle [radius=0.1];
     \draw [fill] (0,0) circle [radius=0.1];
      \draw [fill] (-2,0) circle [radius=0.1];
      \draw [fill] (1,0) circle [radius=0.1];
      \draw [fill] (-1,0) circle [radius=0.1];
      \draw [white] (0,-1.2) circle [radius=0.1];
      \node at (0.5,0.2) {$q$};
      \node at (1.5,0.2) {$q$};
      \node at (-0.5,0.2) {$q$};
      \node at (-1.5,0.2) {$q$};
\end{tikzpicture}
}
 \caption{Three support sets that we use to explore the relation between $\sigma_s^2(\Phi^\bL)/\#(\bL)$ and $q$, and the relation between the noise tolerance of MUSIC and $q$, respectively.}
\label{figsup1}
\end{figure}
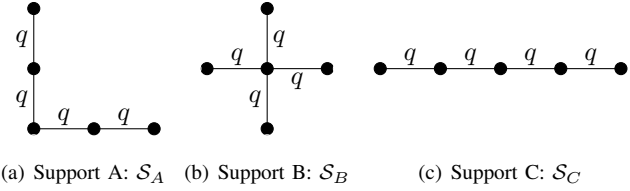

\begin{figure}[hthp]
        \centering
        \subfigure[$\frac{\si_s^2(\Phi^\bL)}{\#(\bL)}$ versus q]{\includegraphics[width=4.35cm]{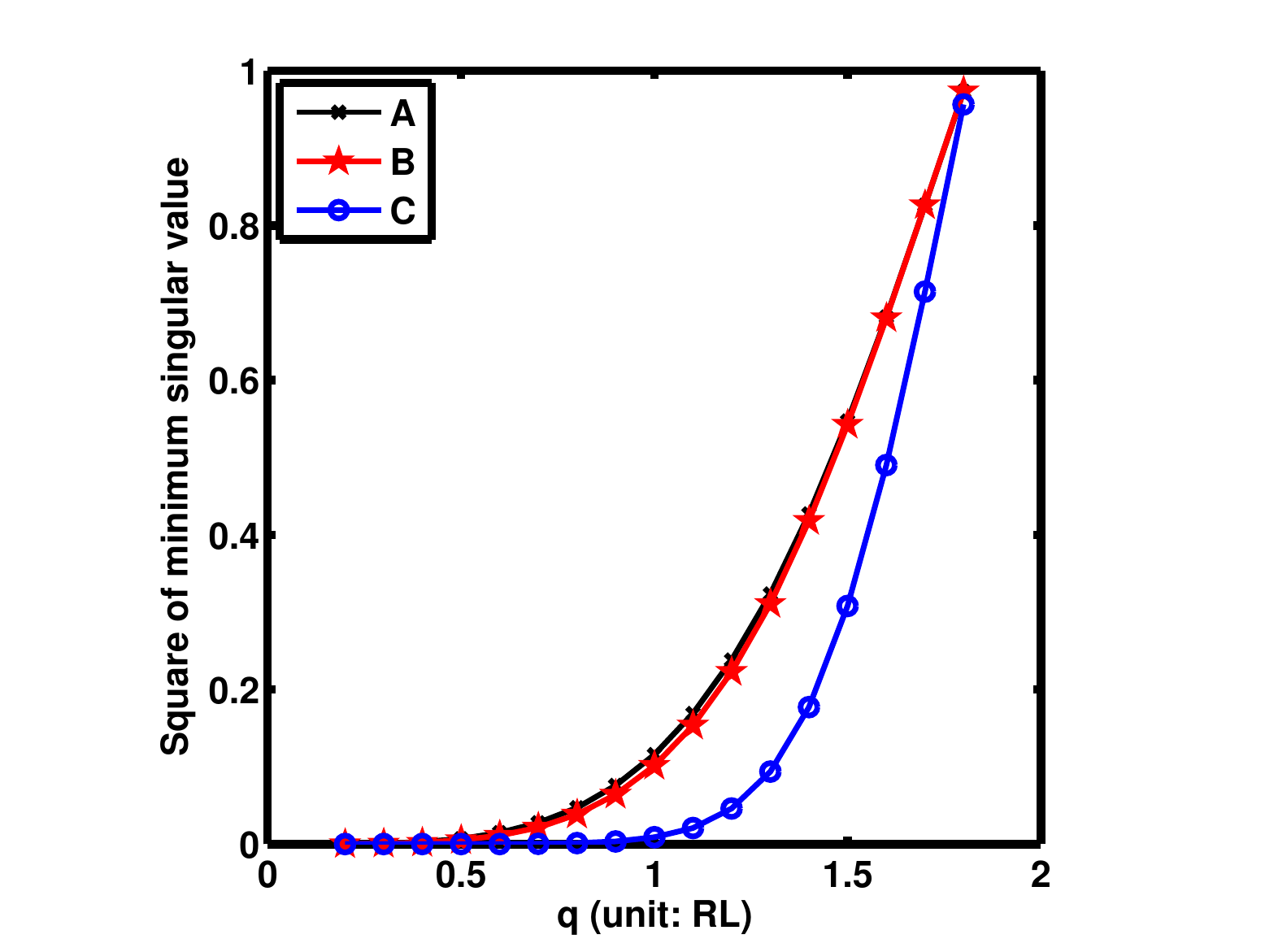}}
      \subfigure[{$\log_{10}[\frac{\si_s^2(\Phi^\bL)}{\#(\bL)}]$ versus $\log_{10}q$}]
      {\includegraphics[width=4.35cm]{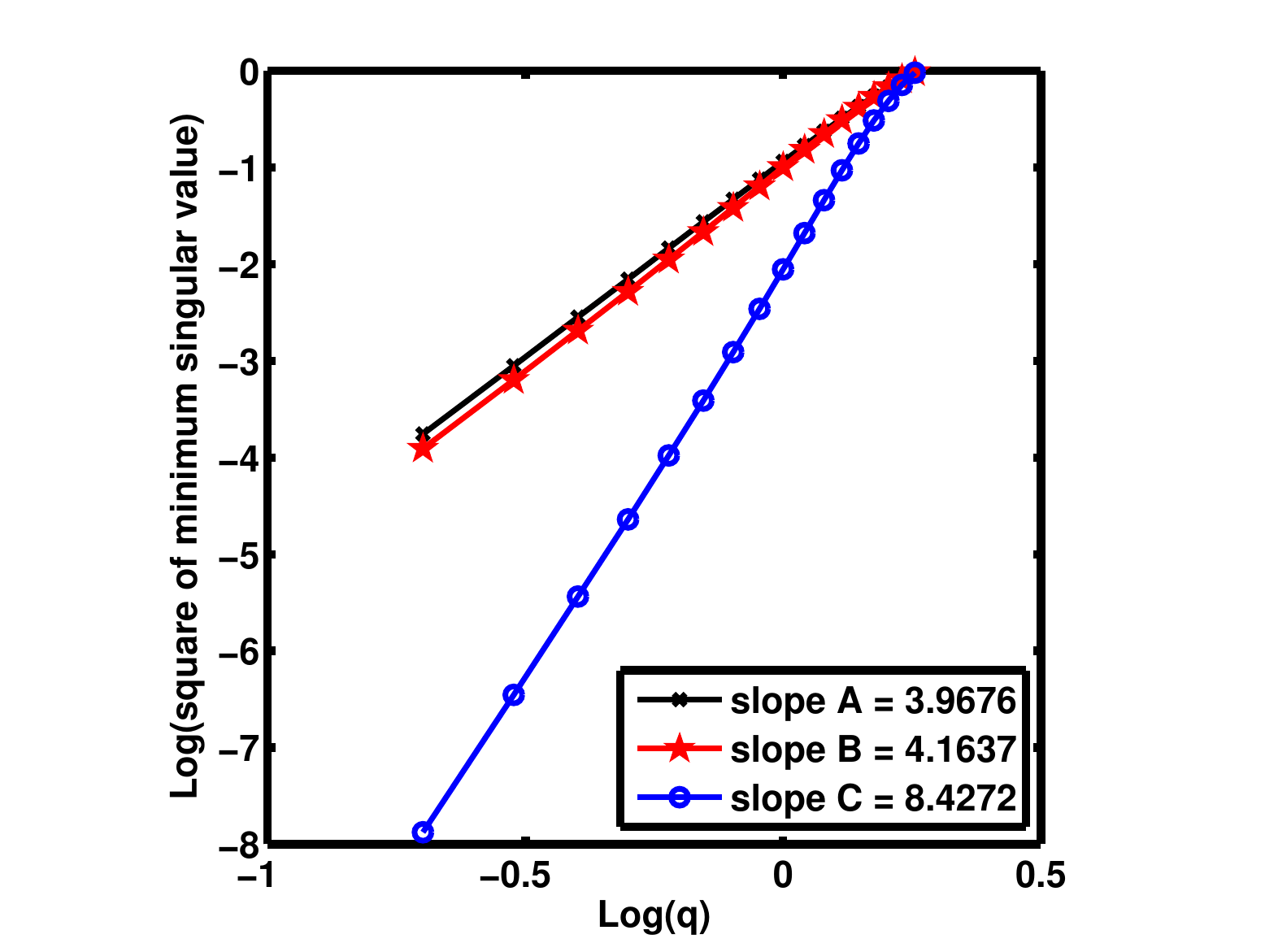}}
        \caption{The plot of $\si_s^2(\Phi^\bL)/\#(\bL)$ versus varied $q$ in ordinary scale (a) and in log-log scale  to the base $10$ (b) for three support sets shown in Figure \ref{figsup1}. It appears that the log-log plots are straight lines so we conclude that, as $q$ decreases, $\si_s^2(\Phi^\bL)/\#(\bL)$ follows a power law with respect to $q$ with a power depending on the support set $\supp$, i.e., $\si_s^2(\Phi^\bL)/\#(\bL) \sim q^{\gamma(\supp)}$. Least squares fitting gives rise to $\gamma(\supp_A) = 3.9676$, $\gamma(\supp_B) = 4.1637$ and $\gamma(\supp_C) = 8.4272$.
  }
     \label{figsup2}
\end{figure}

Let $D = 2$. We consider three support sets containing five frequencies as shown in Figure \ref{figsup1}. For each support set, we fix $\bN = [N_1 \ N_2]^T = [20 \ 20]^T$, one RL = $1/20$ and $\bL = [L_1 \ L_2]^T = [10 \ 10]^T$, construct the two dimensional Vandermonde matrix $\Phi^\bL$ and numerically compute $\si_s^2(\Phi^\bL)/\#(\bL)$. In Figure \ref{figsup2}, for three support sets, the plot of $\si_s^2(\Phi^\bL)/\#(\bL)$ versus varied $q$ is shown in (a) in ordinary scale and in (b) in log-log scale. It appears that the log-log plots are straight lines so we conclude that, as $q$ decreases, $\si_s^2(\Phi^\bL)/\#(\bL)$ follows a power law with respect to $q$ with a power depending on the support set $\supp$, i.e., $\si_s^2(\Phi^\bL)/\#(\bL) \sim q^{\gamma(\supp)}$. Least-square fitting gives rise to $\gamma(\supp_A) = 3.9676$, $\gamma(\supp_B) = 4.1637$ and $\gamma(\supp_C) = 8.4272$.

\begin{figure*}[hthp]
        \centering
     \subfigure[Support A]{\includegraphics[width=5.5cm]{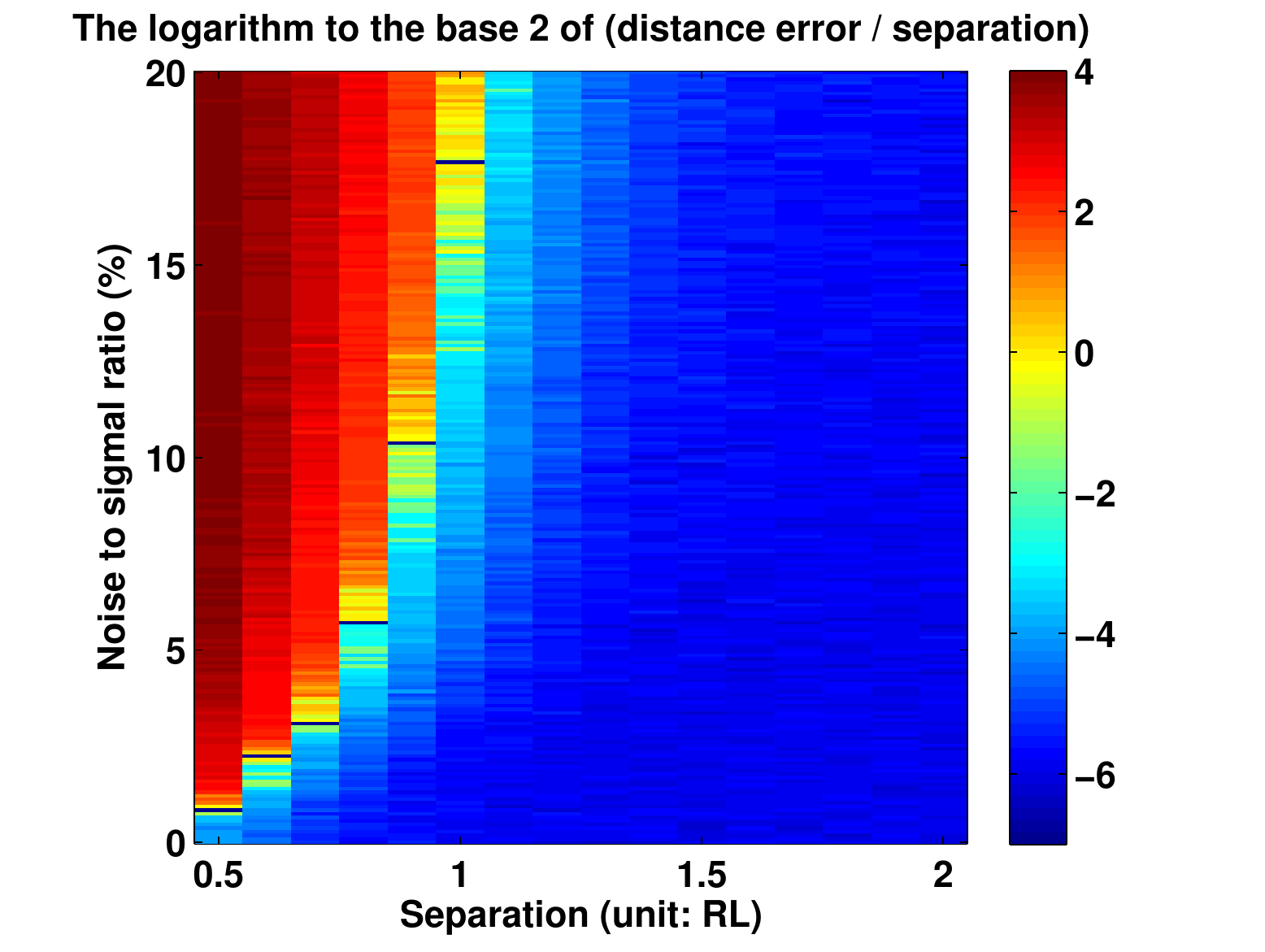}}
      \subfigure[Support B]{\includegraphics[width=5.5cm]{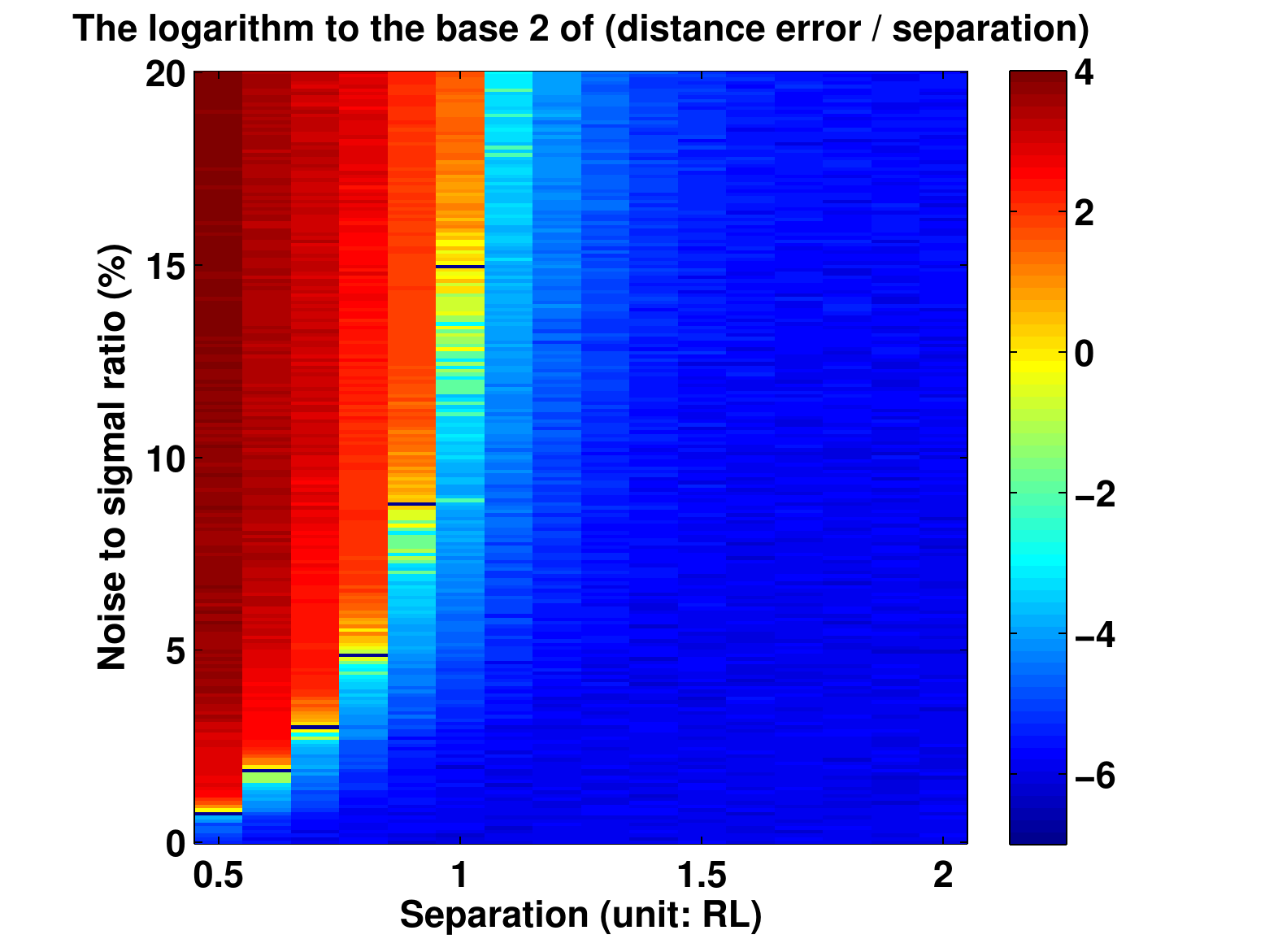}}
      \subfigure[Support C]{\includegraphics[width=5.5cm]{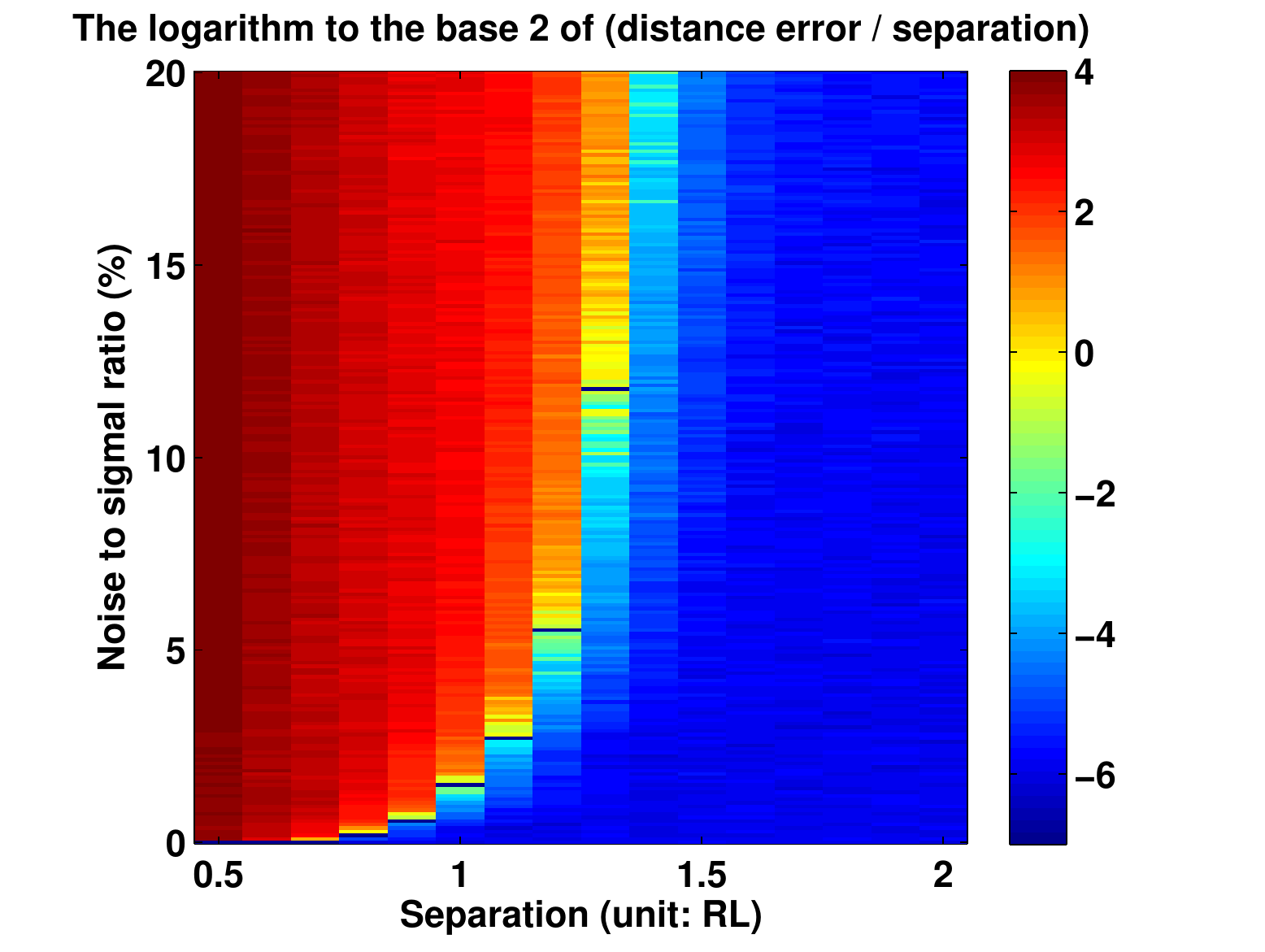}}
      \subfigure[Phase transition curves]{\includegraphics[width=6cm]{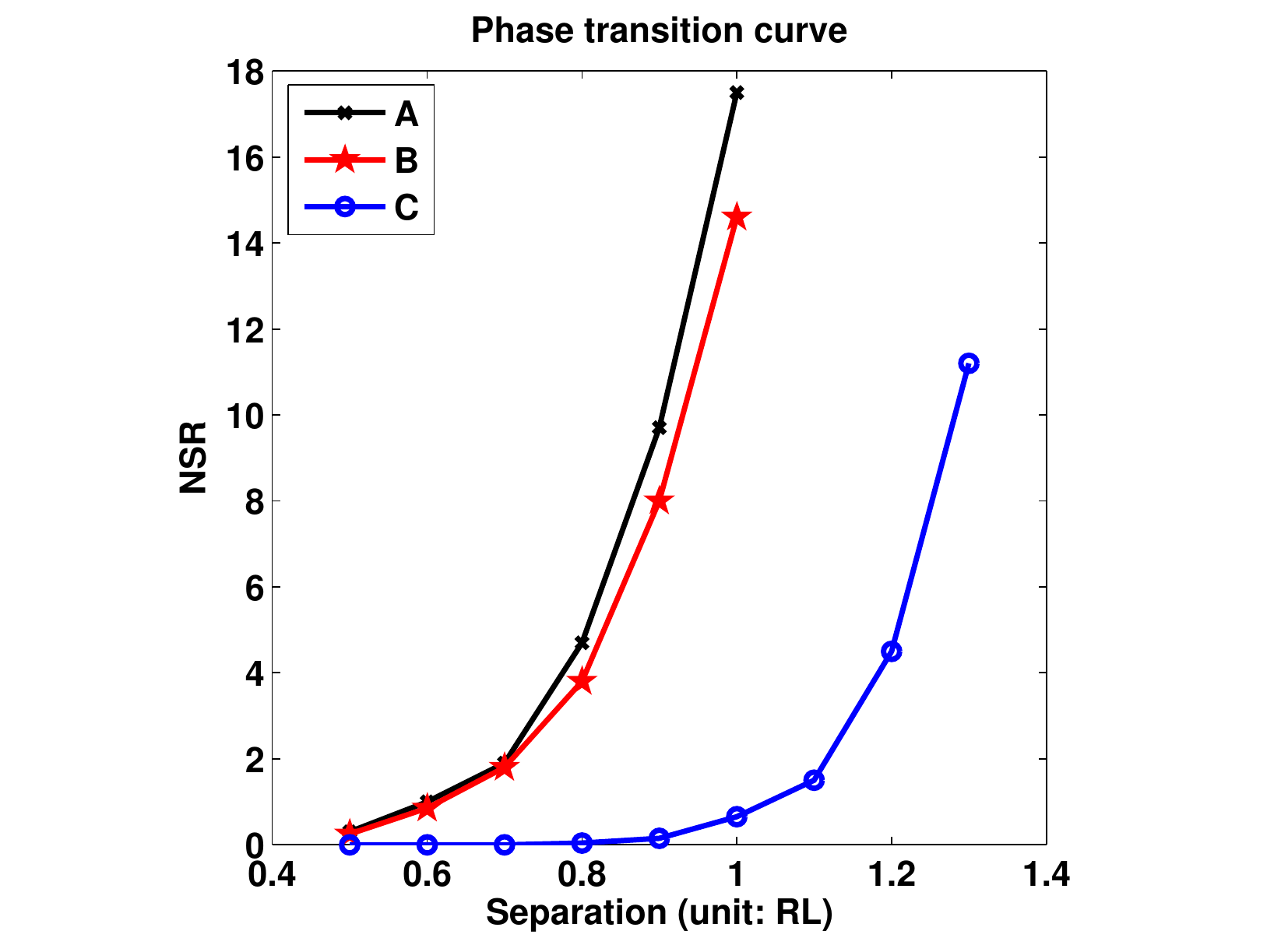}}
      \subfigure[Log-log plot of phase transition curves]{\includegraphics[width=6cm]{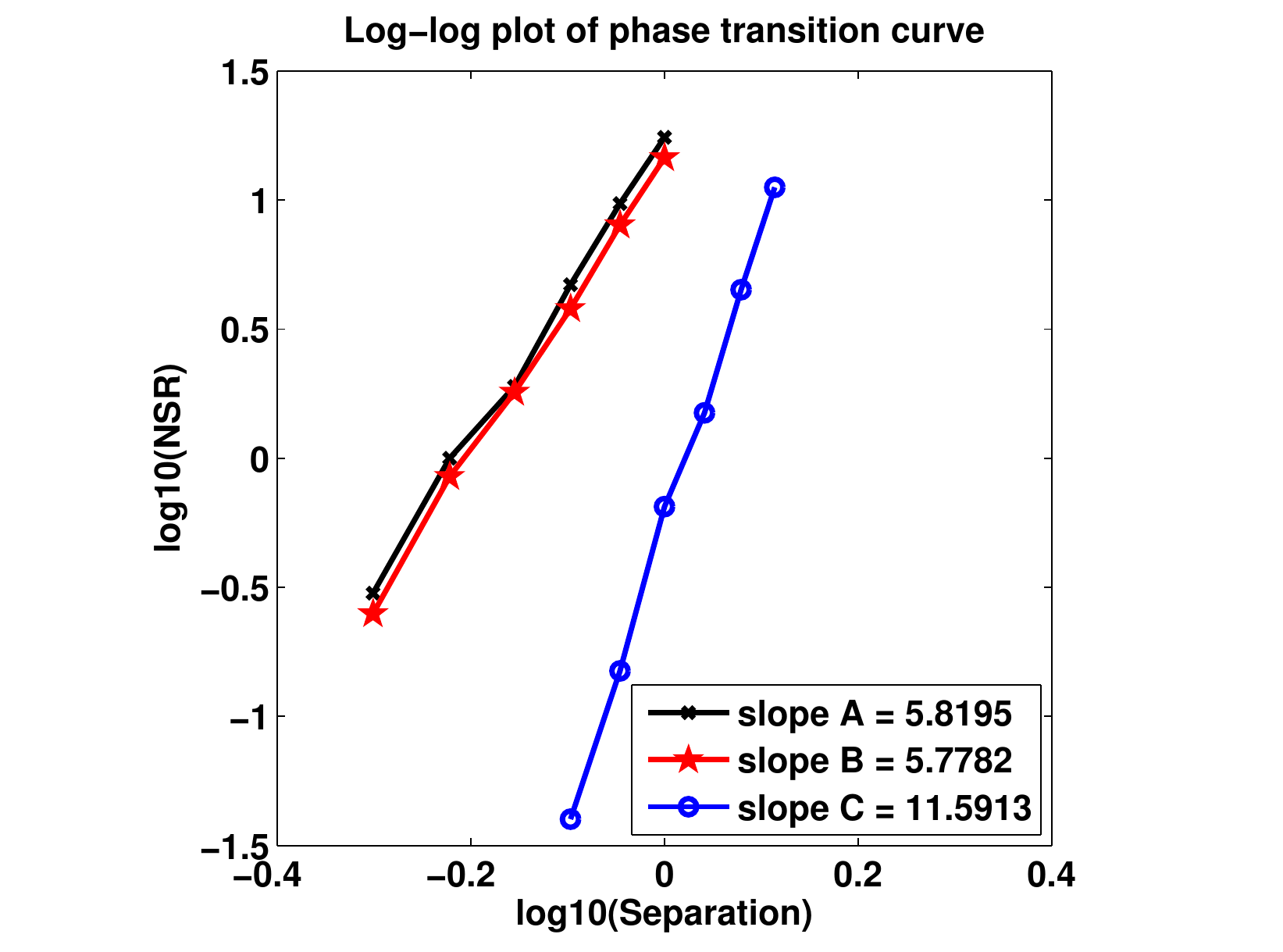}}
     \caption{Color plots in (a)(b)(c) show the logarithm to the base 2 of average $\itd(\supp,\widehat\supp)/q$ with respect to NSR (y-axis) and $q$ (x-axis) in the unit of RL for support A, B and C shown in Fig. \ref{figsup1}, respectively. 
Reconstruction is considered successful if $\itd(\supp,\widehat\supp)/q < 1/2$ (from green to blue). A clear phase transition is observed. Transition points above which $\itd(\supp,\widehat \supp)/q > 1/2$ are marked out by black bars in (a)(b)(c). Phase transition curves are shown in (d) in ordinary scale  and in (e) in log-log scale. 
It appears that the transition curves in the log-log scale can be fitted by straight lines, and therefore, the noise level that MUSIC can tolerate scales like $q^{\tau(\supp)}$ as $q$ decreases. Least squares fitting gives rise to $\tau(\supp_A) = 5.8195$, $\tau(\supp_B) = 5.7782$ and $\tau(\supp_C) = 11.5913$.
  }
     \label{figsup3}
\end{figure*}

Then we run MUSIC on the reconstruction of five randomly phased complex objects supported on the three sets in Figure \ref{figsup1} with varied separation $q$ and varied NSR for 20 trials and record the average of $\itd(\supp,\widehat\supp)/q$. Figure \ref{figsup1} (a)(b)(c) displays the color plot of the logarithm to the base 2 of average $\itd(\supp,\widehat\supp)/q$ with respect to NSR (y-axis) and $q$ (x-axis) in the unit of RL. Frequency localization is considered successful if $\itd(\supp,\widehat\supp)/q <1/2$.  
A  phase transition occurs in (a)(b)(c), manifesting  MUSIC's capability of resolving five closely spaced complex-valued objects supported on $\supp_A$, $\supp_B$ and $\supp_C$ respectively if NSR is below certain level. The phase transition curves above which $\itd(\supp,\widehat \supp)/q > 1/2$ are marked out in black in Figure \ref{figsup3} (a)(b)(c).
These curves are shown in (d) in the ordinary scale  and in (e) in log-log scale to the base $10$. 
It appears that the phase transition curves in the log-log scale can be fitted by straight lines. As a result, the noise tolerance of MUSIC scales like $q^{\tau(\supp)}$ when $q <$ 2 RL. Least-square fitting gives rise to $\tau(\supp_A) = 5.8195$, $\tau(\supp_B) = 5.7782$ and $\tau(\supp_C) = 11.5913$.

\section{Conclusions and extensions}
\label{seccon}
We have extended the performance analysis for MUSIC applied on single-snapshot spectral estimation from one dimension \cite{FLMUSIC} to higher dimensions. 
An explicit estimate on the perturbation of the noise-space correlation function is provided when frequencies are pairwise separated above 2RL at each direction. If the minimum separation among frequencies drops below 2RL, our numerical experiments show that the noise tolerance of MUSIC follows a power law with respect to the minimum separation.

A natural question to ask is: can we apply MUSIC for the reconstruction of sparse frequencies with a few random measurements? The answer is yes!
Performance of MUSIC for the sparse joint frequency estimation problem with compressive measurements was studied in \cite{LiaoSIP}. As for the single-snapshot problem, we combine the Enhanced Matrix Completion (EMC) technique proposed by Chen and Chi in \cite{YY} and MUSIC for a novel strategy of single-snapshot spectral estimation with compressive and noisy measurements. 

In \cite{YY}, a matrix completion technique is used to stably recover $\{y(\bn), \mathbf{0} \le \bn \le \bN\}$ from its partial noisy samples. Let $y$ and $y^e$ be the complete set of noiseless and noisy data as before. Suppose the sampling set $\Lambda$ of size $m$ is uniformly chosen at random  from $\{\bn \ | \ \mathbf{0}\le \bn \le \bN \}$. Let $\calP_\Lambda: \CC^{(N_1+1)\times  \ldots \times (N_D+1)} \rightarrow \CC^{(N_1+1)\times \ldots \times (N_D+1)}$ be the operator of setting the entries outside of $\Lambda$ zero.


Given the noisy data $\{\calP_{\Lambda}  y^e\}$ satisfying  $\|\calP_\Lambda (y^e - y)\|_F \le \delta$, we first apply the following EMC technique proposed in \cite{YY} to recover $y$ from $\calP_\Lambda y^e$:
\begin{align}
 y^{\rm EMC} =  & {\rm arg}\min_{z \in \CC^{(N_1+1)\times  \ldots \times (N_D+1)}} \|{\rm Hankel}(z)\|_*,
 \nonumber
  \\
&
 \text{s.t.}\quad  \|\calP_\Lambda (z-y^e) \|_F \le\delta
 \label{eqyy}
 \end{align}
where $\|\cdot\|_*$ denotes the nuclear norm. 
While EMC fulfills the task of completing the measurements from partial samples, MUSIC can be used for frequency recovery based on $y^{\rm EMC}$.

\commentout{
\begin{table*}
\begin{center}
   \begin{tabular}{|l|}\hline
    { \centerline{\bf Two-dimensional spectral estimation with compressive measurements}} \\ \hline
    {\bf Input:} $\calP_\Lambda y^e,\delta, s, \bL$. \\
    1) Matrix completion:
          ${y^{\rm EMC}} =  {\rm arg}\min_{z \in \CC^{\#(\bN)}} \|{\rm Hankel}(z)\|_*, \  
 \text{subject to } \|\calP_\Lambda (z-y^e) \|_F \le\delta$
    \\
     2) Form Hankel matrix $H^{\rm EMC} = {\rm Hankel}(y^{\rm EMC}) \in \CC^{\#(\bL)\times \#(\bN-\bL)}$.
     \\
     3) SVD: $H^{\rm EMC} = [U^{\rm EMC}_1\  U^{\rm EMC}_2] {\rm diag}({\si}^{\rm EMC}_1 , \ldots , {\si}^{\rm EMC}_s ,\ldots) [{V}^{\rm EMC}_1\ {V}^{\rm EMC}_2]^* $, where ${U}^{\rm EMC}_1 \in \CC^{\#(\bL)\times s}$.\\
     4) Compute imaging function $\CJ^{\rm EMC}(\bom) = \|\phi^{\bL}(\bom)\|_2 /\|{U^{\rm EMC}_2}^* \phi^\bL(\bom)\|_2$. \\
   {\bf Output:} $\supp^{\rm EMC} =\{ \bom \text{ corresponding to the } s \text{ largest local maxima of } {\CJ}^{\rm EMC}(\bom) \} $.\\
    \hline
   \end{tabular}
\end{center}
\caption{EMC $+$ MUSIC for two-dimensional spectral estimation with compressive measurements}
\label{table2}
\end{table*}
}



\appendices

\section{Proof of Theorem \ref{thm2}}
\label{app1}

The proof of Theorem \ref{thm2} relies on the extremal majorant and minorant of the characteristic function $\chi_{\LkLk}$ defined as 
\[ \chi_\LkLk(t) = \left\{
  \begin{array}{l l}
    1 & \quad \text{if $t \in \LkLk$}\\
    0 & \quad \text{otherwise}
  \end{array} \right..\]

\subsection{Extremal majorant and minorant of characteristic functions}

It was observed by A. Beurling \cite{Beurling} that the function
$$B(t) = \left( \frac{\sin \pi t}{\pi}\right)^2
\left\{
\sum_{n =0}^{+\infty} \frac{1}{(t-n)^2}
-\sum_{n =-\infty}^{-1} \frac{1}{(t-n)^2}
+\frac 2 t
\right\}$$
has the following properties:
\begin{enumerate}
\item Its Fourier transform $\widehat{B}(\om) = \int_{-\infty}^{+\infty} B(t)e^{-2\pi i \om t}dt$ is supported on $[-1,1]$.

\item It majorizes sgn($t$) (the signum of $t$): $\sgn(t)\le B(t),$ $\forall t\in \RR$.

\item It satisfies $\int_{-\infty}^{+\infty} \left[
B(t)-\sgn(t)
\right]dt = 1$.

\item It is extremal in the sense that any function $A(t)$ satisfying 1 and 2 follows $\int_{-\infty}^{+\infty} \left[
A(t)-\sgn(t)
\right]dt \ge 1$.
\end{enumerate}

Our proof considers the characteristic function $\chi_\LkLk$ for $ k=1,\ldots,D$ and utilizes its extremal majorant $G^{q_k}_\LkLk$ and minorant $H^{q_k}_\LkLk$ whose Fourier transform is supported on $[-q_k,q_k]$ with $q_k$ given in \eqref{sepding}. 
The extremal majorant $G^{q_k}_\LkLk$ for $\chi_\LkLk$, found by Selbeg in 1974 \cite{Selberg}, coincides with 
$$G^{q_k}_\LkLk(t) = \frac 1 2 \left[ 
B\left(q_k\left(L_k/2-t\right)\right) +B\left(q_k(t+L_k/2)\right)
\right],$$
so that 
\begin{align*}
\chi_\LkLk (t)
& = \frac 1 2
\left[
\sgn\left(q_k(\frac{L_k}{2}-t)\right) + \sgn\left(q_k(t+\frac{L_k}{2})\right)
\right]
\\
&\le G^{q_k}_\LkLk(t).
\end{align*}
One can verify that 
\begin{align*}
& \widehat{G}^{q_k}_\LkLk(\om) = \int_\RR G^{q_k}_\LkLk (t)e^{-2\pi i t \om}dt 
\\
&=  {q_k}^{-1} {e^{-\pi i L_k\om}}\widehat{B}(-\om/q_k) +  q_k^{-1}  e^{\pi i L_k\om}\widehat{B}(\om/q_k),
\end{align*}
so $\widehat{G}^{q_k}_\LkLk$ is supported on $[-q_k,q_k]$.
\begin{proposition}
\label{prop2}
There are functions $H_\LkLk^{q_k}(t)$ and $G_\LkLk^{q_k}(t)$ whose Fourier transform is supported on $[-q_k,q_k]$ and 
$$H_\LkLk^{q_k}(t) \le \chi_\LkLk(t) \le G_\LkLk^{q_k}(t), \forall t\in \RR,$$ 
\begin{align*}
\frac{1}{q_k}&=\int_{-\infty}^{+\infty} \left( G_\LkLk^{q_k}(t)-\chi_\LkLk(t)\right) dt
\\
 &=
\int_{-\infty}^{+\infty} \left( \chi_\LkLk(t)-H_\LkLk^{q_k}(t)\right) dt
.
\end{align*}
\end{proposition}

\begin{figure}
     \centering
     \includegraphics[width=7cm]{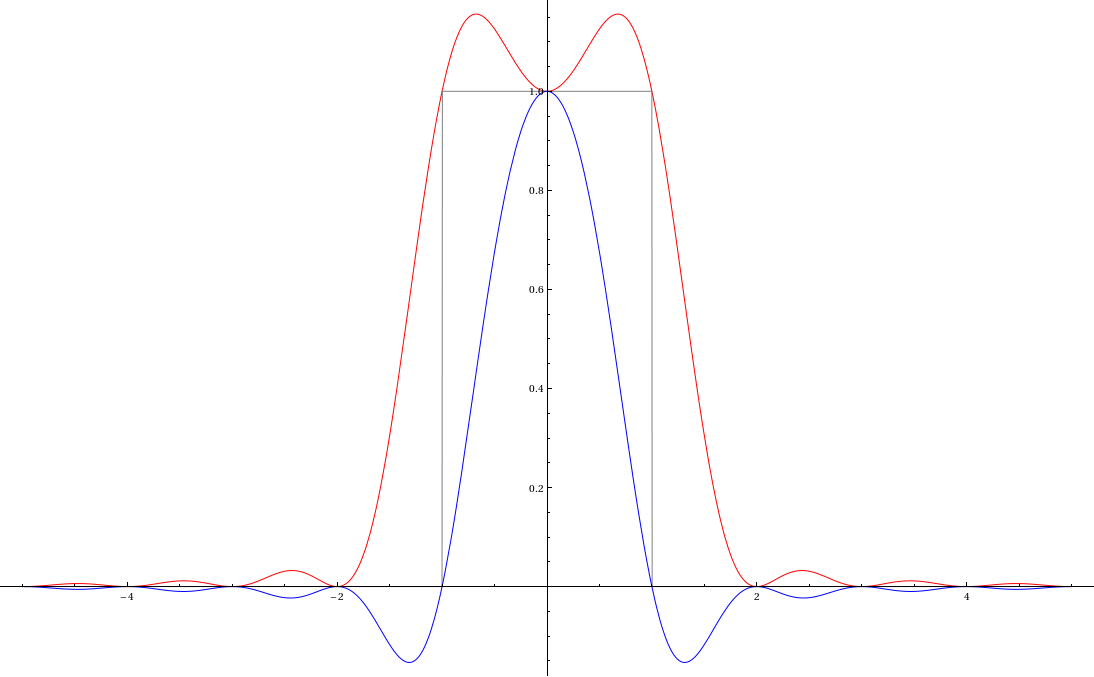}
     \caption{Plot of $G^1_{[-1,1]}$ in red and $H^1_{[-1,1]}$ in blue. 
     }
     \label{figmajorant}
\end{figure}

As an example, $G^1_{[-1,1]}$ and $H^1_{[-1,1]}$ are shown in Figure \ref{figmajorant} in red and blue respectively.

\subsection{Proof of Theorem \ref{thm2}}

\begin{IEEEproof}
We prove Theorem \ref{thm2} for even integers $L_k,\ k=1,\ldots,D$. Let $\bL = (L_1,\ldots,L_D)^T$. Let $\Phi^{-\frac{\bL}{2} \rightarrow \frac{\bL}{2} } = \Phi^\bL \Lambda$
where $\Lambda$ is the $s \times s$ diagonal matrix such that $\Lambda_{jj} = e^{-2\pi i (\frac{L_1}{2} \om^j_1+\frac{L_2}{2}\om^j_2+\ldots+\frac{L_k}{2}\om^j_D)}, \ j=1,\ldots,s$.
$\Phi^{-\frac\bL 2 \rightarrow \frac\bL 2}$ and $\Phi^\bL$ share the same singular values, so it suffices to prove \eqref{ding} for $\Phi^{-\frac\bL 2 \rightarrow \frac\bL 2}$.

Denote the $D$ dimensional rectangular region $[-\frac{L_1}{2},\frac{L_1}{2}] \times [-\frac{L_2}{2},\frac{L_2}{2}] \times \ldots \times [-\frac{L_D}{2},\frac{L_D}{2}]$ by $\LL$. Let $\bq = (q_1,\ldots,q_D)^T$.
We use the following majorant for the characteristic function $\chi_\LL$
$$\bG^\bq_\LL(\bt) = \prod_{k=1}^D G^{q_k}_{\LkLk}(t_k)$$
to prove the upper bound.
\begin{align}
& \|\Phi^{-\frac\bL 2 \rightarrow \frac\bL 2} \bc\|^2
 = \sum_{{ \bn \in \ZZ^D }}
\chi_{\LL}(\bn) \left|\sum_{j=1}^s c_j e^{2\pi i \bom^j \cdot \bn}\right|^2 
\nonumber
\\
&\le  
\sum_{\substack{ \bn \in \ZZ^D }} \bG^\bq_\LL(\bn) \left|\sum_{j=1}^s c_j e^{2\pi i \bom^j \cdot \bn}\right|^2 
\nonumber
\\
&
= \sum_{\substack{ \bn \in \ZZ^D }}\bG^\bq_\LL(\bn) 
\left(\sum_{j=1}^s  \bar{c}_j e^{-2\pi i \bom^j \cdot \bn}  \right) 
\left(\sum_{l=1}^s  c_l e^{2\pi i \bom^l \cdot \bn}  \right) 
\nonumber
\\
&
=
\sum_{j=1}^s
\sum_{l=1}^s  
\bar{c}_j c_l 
\sum_{\substack{ \bn \in \ZZ^D }} \bG^\bq_\LL(\bn)
e^{-2\pi i (\bom^j-\bom^l) \cdot \bn}
\label{thm2eq1}
\\
&=
\sum_{j=1}^s
\sum_{l=1}^s  
\bar{c}_j c_l 
\sum_{\br\in\ZZ^D}\widehat{\bG}^\bq_\LL(\bom^j-\bom^l-\br),
\label{thm2eq2}
\end{align}
where 
\begin{align*}
& \widehat\bG^\bq_\LL(\bom)
 = \int_{\RR^D} \bG^\bq_\LL(\bt)e^{-2\pi i \bom\cdot \bt}d\bt
\\
& 
= \prod_{k=1}^D \int_\RR G^{q_k}_\LkLk(t_k) e^{-2\pi i \om_k t_k}dt_k 
= \prod_{k=1}^D \widehat{G}^{q_k}_\LkLk(\om_k).
\end{align*}
and the equality from \eqref{thm2eq1} to \eqref{thm2eq2} is from Poisson summation formula. According to Proposition \ref{prop2}, $\widehat{G}^{q_k}_\LkLk$ is supported on $[-q_k, q_k]$. For any $\bom^j,\bom^l \in \supp, j\neq l$, we assume $\itd_k(\bom^j,\bom^l) \ge q_k$ for $k=1,\ldots,D$. Therefore $\sum_{\br\in\ZZ^D}\widehat{\bG}^\bq_\LL(\bom^j-\bom^l-\br) = 0$ if $j \neq l$ and then
\begin{align}
&\|\Phi^{-\frac\bL 2 \rightarrow \frac\bL 2} \bc\|^2
 \le
 \|\bc\|^2 \widehat{\bG}^\bq_\LL(\mathbf{0})
 \nonumber
 \\
 &= \|\bc\|^2 \prod_{k=1}^D
 \left(\widehat{G}^{q_k}_\LkLk(0)\right)
 \nonumber
 \\
 & =\|\bc\|^2 
  \prod_{k=1}^D \left(\int_{-\infty}^{+\infty} G^{q_k}_\LkLk(t)dt\right)
  \nonumber
  \\
& =\|\bc\|^2 
\prod_{k=1}^D \left(\int_{-\infty}^{+\infty} \chi_\LkLk(t)dt+\frac{1}{q_k}\right) 
\\
&
= \|\bc\|^2 \prod_{k=1}^D\left(L_k+\frac{1}{q_k}\right).
\label{thm2eq3}
 \end{align}
 
 The lower bound is proved in a similar way by using the following minorant for $\chi_\LL$:
 $$\bH^\bq_\LL(\bt) = \prod_{k=1}^D H^{q_k}_\LLD(t_k).$$
\begin{align}
& \|\Phi^{-\frac\bL 2 \rightarrow \frac\bL 2} \bc\|^2
\ge
\sum_{j=1}^s
\sum_{l=1}^s  
\bar{c}_j c_l 
\sum_{\br\in\ZZ^D}\widehat{\bH}^\bq_\LL(\bom^j-\bom^l-\br)
\nonumber
\\
& = \|\bc\|^2 \widehat{\bH}^\bq_\LL(\mathbf{0})
= \|\bc\|^2 \prod_{k=1}^D \left( L_k-\frac{1}{q_k}
\right).
\label{thm2eq4}
\end{align}

Combing \eqref{thm2eq3} and \eqref{thm2eq4} gives rise to \eqref{ding}.

\end{IEEEproof}

\section{Proof of Theorem \ref{thm4}}
\label{app3}

\begin{IEEEproof}

According to Theorem \ref{thm2}, 
\small{
\begin{align}
\si_1(H)  
&\le \xmax\smax(\Phi^\bL)\smax(\Phi^{\bN-\bL})
\nonumber
 \\
 & \le \xmax \sqrt{\#(\bL-\mathbf 1)\#(\bN-\bL-\mathbf 1)}
 \nonumber
 \\
 &
\cdot \sqrt{
 \prod_{k=1}^D \left(1+\frac{1}{q_k L_k} \right)
 \left(1+\frac{1}{q_k(N_k-L_k)} \right)
 }
 \label{app3e1}
\end{align}
and
\begin{align}
\si_s(H)  
&\ge \xmin\smin(\Phi^\bL)\smin(\Phi^{\bN-\bL})
\nonumber
\\
& \ge \xmin \sqrt{\#(\bL-\mathbf 1)\#(\bN-\bL-\mathbf 1)}
\nonumber
 \\
 &\cdot \sqrt{
 \prod_{k=1}^D \left(1-\frac{1}{q_k L_k} \right)
 \left(1-\frac{1}{q_k(N_k-L_k)} \right)
 }
 \label{app3e2}
\end{align}
}

Overall \eqref{app3e1}, \eqref{app3e2} and the estimate in Lemma \ref{lemmap1} give rise to \eqref{thm4e2}.
\end{IEEEproof}

\section{Proof of Theorem \ref{thm5}}
\label{app2}

Proof of Theorem \ref{thm5} is based on the following inequality.

\begin{proposition}[Corollary 4.2.1 \cite{tropp}]
\label{proprm}
Consider a finite sequence $\{Z_k\}$ of fixed complex matrices with dimension $m_1 \times m_2$, and let $\{\gamma_k\}$ be a finite sequence of independent standard normal variables. Form the matrix Gaussian series 
$$Z = \sum_k \gamma_k Z_k.$$
Compute the variance parameter
$$\nu^2 = \nu^2(Z) = \max
\{
\|\EE(Z Z^*)\|_2,
\|\EE(Z^* Z)\|_2
\}.$$
Then
$$\EE \|Z\|_2 \le \sqrt{2\nu^2 \log(m_1+m_2)}$$
and furthermore, for all $t \ge 0$,
$$\PP\{\|Z\|_2 \ge t\} \le (m_1+m_2)\exp\left(-\frac{t^2}{2\nu^2}\right).$$
\end{proposition}

Here is the proof of Theorem \ref{thm5}.
\begin{IEEEproof}
Recall that $E = {\rm Hankel}(e)$ where the Hankel matrix is formed according to Section \ref{sechankel}. Let $B^{n_1,\ldots,n_D}$ be the ${\#(\bL) \times \#(\bN-\bL)}$ matrix whose entries are $0$ except those corresponding to the locations of $e(n_1,\ldots,n_D)$ in $E$ being $1$. Then
\beq
E = \sum_{n_1 = 0}^{N_1} \ldots \sum_{n_D = 0}^{N_D}
e(n_1,\ldots,n_D) B^{n_1,\ldots,n_D}.
\eeq
Since $e(n_1,\ldots,n_D)$ are independent normal random variables, we can apply Proposition \ref{proprm} to estimate $\|E\|_2$.
All we need is the variance parameter $\nu^2(E)$. On the one hand,
{\small
\begin{align}
& \|\EE(E E^*)\|_2 
\nonumber
 \\
 & = 
\left\|
\sum_{n_1=0}^{N_1} \ldots \sum_{n_D = 0}^{N_D}
\EE 
e^2(n_1,\ldots,n_D)
B^{n_1,\ldots,n_D} (B^{n_1,\ldots,n_D})^T
\right\|_2
\nonumber
\\
& = \left\| \sigma^2  I_{\#(\bL) \times \#(\bL)}\prod_{k=1}^D (N_k-L_k+1)\right\|_2
\nonumber
\\&
= \sigma^2 \prod_{k=1}^D (N_k-L_k+1)
= \sigma^2 \#(\bN-\bL),
\label{app2e1}
\end{align}
}
where $I_{\#(\bL) \times \#(\bL)}$ is the identity matrix of size $\#(\bL) \times \#(\bL)$.

On the other hand,
\begin{align}
& \|\EE(E^* E)\|_2
\nonumber
\\
& 
= 
\left\|
\sum_{n_1=0}^{N_1} \ldots \sum_{n_D = 0}^{N_D}
\EE 
e^2(n_1,\ldots,n_D)
(B^{n_1,\ldots,n_D})^T B^{n_1,\ldots,n_D}
\right\|_2
\nonumber
\\
& = \left\| \sigma^2  I_{\#(\bN-\bL) \times \#(\bN-\bL)}\prod_{k=1}^D (L_k+1)\right\|_2
\nonumber
\\
&
= \sigma^2 \prod_{k=1}^D (L_k+1)
= \sigma^2 \#(\bL).
\label{app2e2}
\end{align}

Combining \eqref{app2e1} and \eqref{app2e2} yields
$$\nu^2 = \nu^2(E) = \sigma^2 \max\{ \#(\bL),\#(\bN-\bL)\}.$$
Then 
{\small
$$
\EE \|E\|_2 \le  \sigma\sqrt{2 \max\{ \#(\bL),\#(\bN-\bL)\} \log(\#(\bL)+\#(\bN-\bL))}
$$
and
} furthermore, for all $t \ge 0$,
{\small
\begin{align*}
&\PP\{\|E\|_2 \ge t\} \le 
\\
& (\#(\bL)+\#(\bN-\bL))\exp
\left(-\frac{t^2}{2\sigma^2 \max\{\#(\bL),\#(\bN-\bL)\}}\right).
\end{align*}
}
\end{IEEEproof}

\section*{Acknowledgment}

The author would like to thank Albert Fannjiang for inspiring discussions, the authors in \cite{Xu2D} for providing their code and anonymous referees for their valuable comments.

\ifCLASSOPTIONcaptionsoff
  \newpage
\fi



%

%

\begin{IEEEbiography}[{\includegraphics[width=1in,height=1.25in,clip,keepaspectratio]{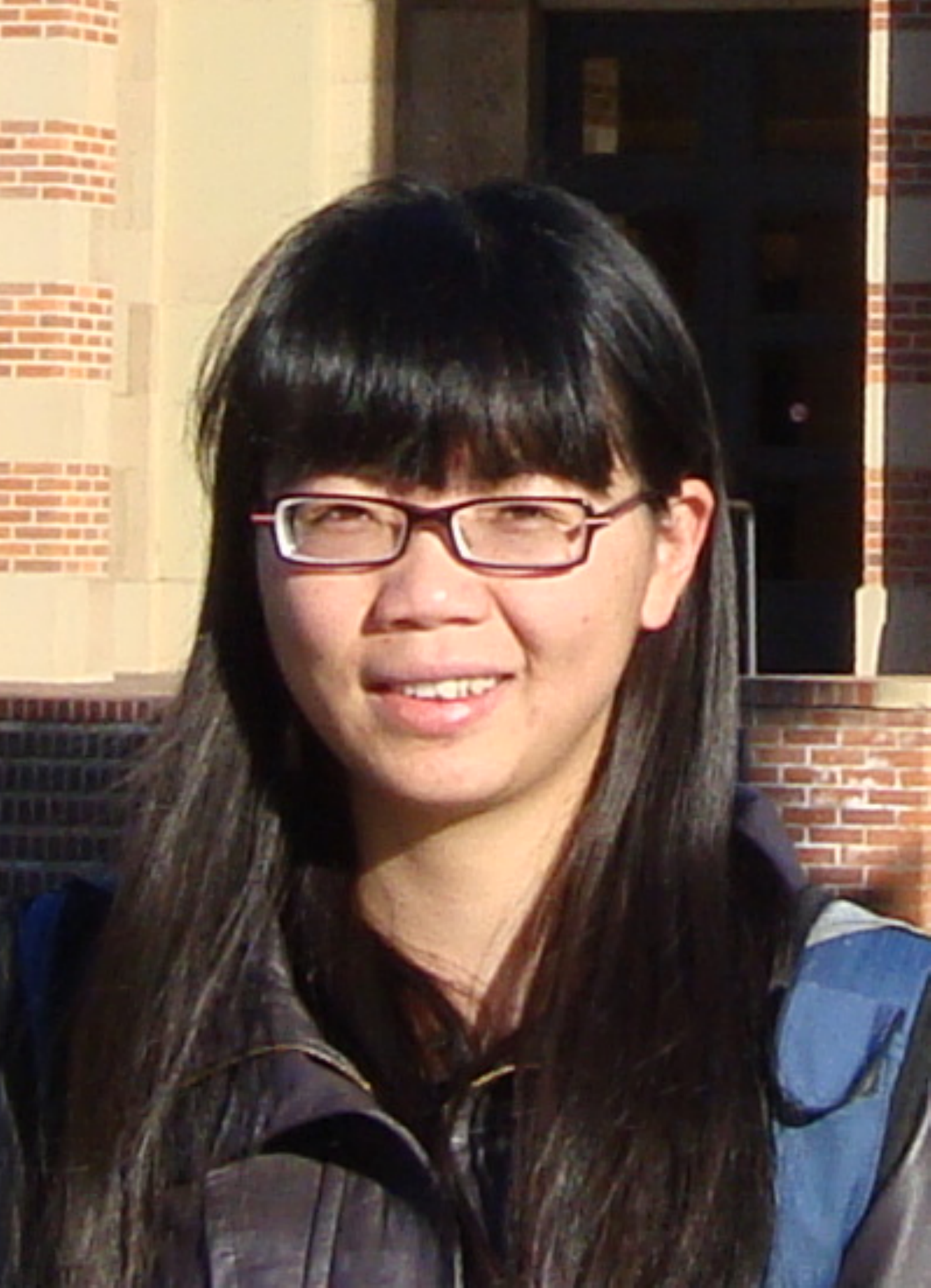}}]{Wenjing Liao}
received the B.Sc. degree in Mathematics from Fudan University of China in 2008 and the Ph.D. in Applied Mathematics from University of California, Davis, CA. She is currently a postdoctoral scholar with the Department of Mathematics at Duke University, Durham, NC and Statistical and Applied Mathematical Sciences Institute, Durham, NC. Her research interests include imaging, signal processing, applied harmonic analysis and machine learning.
\end{IEEEbiography}





\end{document}